 \pdfoutput=1
\documentclass[amsmath,amssymb,preprintnumbers,nofootinbib,a4paper,11pt]{article}
 
\usepackage{graphicx}
\usepackage{epstopdf}
\usepackage{jheppub}
\usepackage{multirow}               
\usepackage{amsmath,amsthm,latexsym,amssymb,amsfonts,epsfig}
\usepackage{fontenc,layout}
\usepackage{hyperref}
\usepackage{psfrag}
\usepackage[cp1250]{inputenc}
\usepackage[usenames,dvipsnames]{xcolor} 
 
\usepackage{array}
\newcommand{\PreserveBackslash}[1]{\let\temp=\\#1\let\\=\temp}
\newcolumntype{C}[1]{>{\PreserveBackslash\centering}p{#1}}
\newcolumntype{R}[1]{>{\PreserveBackslash\raggedleft}p{#1}}
\newcolumntype{L}[1]{>{\PreserveBackslash\raggedright}p{#1}}

\newcommand{\mhone}{m_{H_1}}
\newcommand{\mhtwo}{m_{H_2}}

\newcommand{\be}{\begin{equation}}
\newcommand{\ee}{\end{equation}}

\def\bsp#1\esp{\begin{split}#1\end{split}}
\def\bpm{\begin{pmatrix}}
\def\epm{\end{pmatrix}}

\newcommand{\bea}{\begin{eqnarray}}  
\newcommand{\eea}{\end{eqnarray}}  
 \def\bsp#1\esp{\begin{split}#1\end{split}}
 
\title{Enlarging the scope of resonant di-Higgs searches: \\
Hunting for Higgs-to-Higgs cascades in $4b$ final states \\
at the LHC and future colliders
}

\preprint{\begin{flushright}
TTP19-030 \\ IFT-UAM/CSIC-19-132 \\ CP3-19-49
\end{flushright}}

\author[1]{D. Barducci,}
\author[2]{K. Mimasu,}
\author[3,4]{J.~M.~No,}
\author[5]{C. Vernieri}
\author[6,7]{and J. Zurita}

\affiliation[1]{Dipartimento di Fisica Universit\`a degli Studi di Roma La Sapienza and INFN Sezione di Roma, Piazzale Aldo Moro 5, 00185, Roma, Italy}
\affiliation[2]{Centre for Cosmology, Particle Physics and Phenomenology (CP3), Universite
catholique de Louvain, Chemin du Cyclotron, 2, B-1348 Louvain-la-Neuve, Belgium}
\affiliation[3]{Departamento de F\'isica Te\'orica, Universidad Aut\'onoma de Madrid, 28049, Madrid, Spain}
\affiliation[4]{Instituto de F\'isica Te\'orica, IFT-UAM/CSIC,
Cantoblanco, 28049, Madrid, Spain}
\affiliation[5]{SLAC National Accelerator Laboratory, 2575 Sand Hill Road, Menlo Park, CA 94025-7090, USA}
\affiliation[6]{Institute for Nuclear Physics (IKP), Karlsruhe Institute of Technology,
Hermann-von-Helmholtz-Platz 1, D-76344 Eggenstein-Leopoldshafen, Germany}
\affiliation[7]{Institute for Theoretical Particle Physics (TTP), Karlsruhe Institute of Technology,
Engesserstra\ss{}e 7, D-76128 Karlsruhe, Germany}

\abstract{We extend the coverage of resonant di-Higgs searches in the $b \bar{b} b \bar{b}$ final state to the process $p p \to H_1 \to H_2 H_2 \to b \bar{b} b \bar{b}$, where both $H_{1,2}$ are spin-$0$ states beyond the Standard Model. Such a process constitutes a joint discovery mode for the new states $H_1$ and $H_2$. We present the first sensitivity study of this channel, using public LHC data to validate our analysis.
We also provide a first estimate of the sensitivity of the search for the HL-LHC and future facilities like the HE-LHC and FCC-hh. 
We analyze the discovery potential of this search for several non-minimal scalar sector scenarios: an extension of the SM with two extra singlet scalar fields, the two-Higgs-doublet model and a two-Higgs doublet model plus a singlet, which captures the scalar potential features of the NMSSM. We find that this channel represents a novel, very powerful probe for extended Higgs sectors, offering complementary sensitivity to existing analyses.}

\notoc

\begin{document}

\maketitle

\newpage

\tableofcontents

\vspace{1cm}

\section{Introduction}
\label{sec:intro}

While analyses at the Large Hadron Collider (LHC) by ATLAS and CMS show that the properties of the Higgs particle $h$ with mass $m_h\sim$~125~GeV
are, at present, compatible with those of the Standard Model (SM) Higgs boson $h_{\mathrm{SM}}$~\cite{TheATLASandCMSCollaborations:2015bln,Sirunyan:2018koj,ATLAS-CONF-2019-005},
the detailed nature of the scalar sector responsible for electroweak (EW) symmetry-breaking remains to be determined. It is particularly important to ascertain whether the scalar sector consists of just one $\mathrm{SU(2)}_L$ doublet or has a richer structure with additional states. Addressing this question is a key goal of present and future studies at the LHC.

Searching for the existence of additional Higgs bosons at the LHC constitutes the main avenue for probing non-minimal scalar sectors, allowing to directly access the spectrum and properties of the scalars beyond the SM (BSM). Among such direct searches, those targeting decay chains involving several scalar states (henceforth {\sl Higgs-to-Higgs}) are of particular importance. They depend on the scalar self-couplings and could therefore provide insight into the structure of the scalar potential. 
Resonant di-Higgs production, $p p \to H \to h_{\mathrm{SM}} h_{\mathrm{SM}}$, is the prime (and simplest) example of a Higgs-to-Higgs process, where a resonantly produced BSM $H$ state decays into a pair of $125\;$GeV Higgs bosons $h_{\mathrm{SM}}$ (see~\cite{DiMicco:2019ngk} for a review). 
ATLAS and CMS have looked for this process at $\sqrt{s} = 8$ TeV and $13$ TeV in a wide range of final states, including $b\bar{b} b\bar{b}$~\cite{Khachatryan:2015yea,Aad:2015uka,Sirunyan:2018zkk,Aaboud:2018knk}, 
$b\bar{b} W^+ W^-$~\cite{Sirunyan:2017guj,Aad:2015xja}, $b \bar{b} \tau^+ \tau^-$~\cite{Sirunyan:2017tqo,Sirunyan:2017djm,Aaboud:2018sfw} 
and $b\bar{b} \gamma \gamma$~\cite{Aad:2014yja,Khachatryan:2016sey,Sirunyan:2018iwt,Aaboud:2018ftw}.

Non-minimal Higgs sectors generically feature several BSM states. In such a case, Higgs-to-Higgs decays with both the parent particle and its decay products as BSM states are possible, and may constitute the most promising avenue for their discovery. This has been emphasized in the literature for certain processes within the two-Higgs-doublet model (2HDM)~\cite{Coleppa:2014hxa,Dorsch:2014qja,Haber:2015pua,Dorsch:2016tab,Kling:2016opi}, the 2HDM plus a scalar singlet~\cite{Baum:2018zhf}, the next-to-minimal supersymmetric Standard Model (NMSSM)~\cite{Barducci:2015zna,Aggleton:2016tdd,Baum:2017gbj,Ellwanger:2017skc,Baum:2019pqc} and the SM extended by several singlet scalars~\cite{Robens:2019kga}.

In this work we show that it is possible to enlarge the scope of resonant di-Higgs $p p \to H \to h_{\mathrm{SM}} h_{\mathrm{SM}}$ searches to probe more general Higgs-to-Higgs processes involving two BSM states. We present a detailed sensitivity study of the 
channel $p p \to H_1 \to H_2 H_2 \to b \bar{b} b \bar{b}$ where the heavier state $H_1$ is assumed to be produced via gluon fusion with a subsequent on-shell decay into a pair of $H_2$ bosons ({\emph{i.e}} we consider $m_{H_1}>2 m_{H_2}$).
The potentially dominant $H_1 \to H_2 H_2$ branching fraction for $m_{H_1} \gg m_{H_2}$ combined with a large $H_2 \to b\bar{b}$ branching 
fraction\footnote{A concrete example of such a scenario would be a 2HDM with a large mass splitting $m_{H} \gg m_{A}$ between the CP-even ($H$) and CP-odd ($A$) neutral BSM scalars.} typical of light scalars (which would at the same time make the discovery of $H_2$ via direct production challenging, see {\emph{e.g.}}~\cite{Vernieri:2014wfa})
make this search channel an important, yet unexplored, probe of non-minimal Higgs sectors.  
While no ATLAS or CMS analysis of the $ p p \to H_1 \to H_2 H_2 \to b \bar{b} b \bar{b}$ signature 
exists at present\footnote{We note there are 
existing LHC analyses for $h_{\mathrm{SM}} \to H_2 H_2$, with $m_{H_2} < 62$ GeV, see {\emph{e.g.}}~\cite{Khachatryan:2017mnf,Aaboud:2016oyb}.}, we can use its similarity to resonant di-Higgs searches in the $b \bar{b} b \bar{b}$ final state to validate our analysis for $m_{H_2} = 125$ GeV, before extending it to the 2D mass parameter space ($m_{H_1},\,m_{H_2}$). 
Specifically, we follow the recent $\sqrt{s} = 13$ TeV CMS search for a narrow spin-$0$ or spin-$2$ di-Higgs resonance in the $b \bar{b} b \bar{b}$ final state 
with $35.9$ fb$^{-1}$ of integrated luminosity~\cite{Sirunyan:2018zkk}. The detailed  public information available for this search allows us to reproduce their reported selection efficiencies and 95\% confidence level (C.L.) exclusion sensitivities with our simulation. We then obtain the expected signal efficiencies for the $p p \to H_1 \to H_2 H_2 \to b \bar{b} b \bar{b}$ process in the mass plane ($m_{H_1},\,m_{H_2}$) and provide the 13 TeV LHC 95\% C.L. exclusion sensitivity on the signal cross section with $35.9$ fb$^{-1}$, as a function of $m_{H_1}$ and $m_{H_2}$.

In addition to the above, we provide an extrapolation of the current exclusion sensitivity for the
High-Luminosity (HL)-LHC with $\sqrt{s} = 14$ TeV, as well as to future collider proposals like the High-Energy (HE)-LHC with $\sqrt{s} = 27$ TeV and a $\sqrt{s} = 100$ TeV proton-proton collider (henceforth referred to as FCC-hh). 
We discuss the impact of multi-jet background systematic uncertainties on the expected sensitivity of our proposed search, as well as the effect of possible future improvements on the $b$-tagging and trigger efficiencies.

Finally, we assess the reach of the proposed analysis within specific BSM models: a two-singlet extension of the SM, a 2HDM scenario and a 2HDM plus a real scalar or pseudoscalar singlet. This allows us to compare the projected sensitivity of the search to other analyses targeting BSM scalars, studying their complementarity and identifying where the search $p p \to H_1 \to H_2 H_2 \to b \bar{b} b \bar{b}$ provides the leading probe of the existence of the $H_1$ and $H_2$ states.

\vspace{1mm}
Our work is organised as follows: in section~\ref{sec:validation} we reproduce the efficiencies of the CMS $\sqrt{s} = 13$ TeV resonant di-Higgs search in the $b \bar{b} b \bar{b}$ final state, in order to validate our subsequent analysis.
In section~\ref{sec:enlarging} we derive the present 95\% C.L. exclusion sensitivity prospects for the $ p p \to H_1 \to H_2 H_2 \to b \bar{b} b \bar{b}$ process, 
and in section~\ref{sec:enlarging_2} we provide extrapolations to the HL-LHC, the HE-LHC and the FCC-hh. In section~\ref{sec:models} we cast these prospects into 
a two-singlet extension of the SM, a 2HDM scenario and a 2HDM $+$ singlet scalar/pseudoscalar, comparing in each case the sensitivity of our proposed search with other LHC searches for BSM scalars. 
Finally, we summarize our results in section~\ref{sec:conclusions}. 


\section{Implementation and validation of the CMS search }
\label{sec:validation}

As stated in section~\ref{sec:intro}, there is no current experimental search at the LHC for BSM spin-$0$ states $H_{1,2}$ ({\emph{i.e.}}~belonging to an extended Higgs sector) through the process $p p \to H_1 \to H_2 H_2$. However, the process bears a strong similarity to resonant di-Higgs production 
$p p \to H_1 \to h_{\rm SM} h_{\rm SM}$, which has been actively searched for by ATLAS and CMS since LHC Run 1.
This similarity can thus be exploited to extend current resonant di-Higgs searches to include processes where both scalars belong to a BSM sector. In analogy with resonant di-Higgs, different analysis strategies can be devised, depending on the decay channels of the $H_2$ scalar. In this work, we concentrate on the $b \bar{b} b \bar{b}$ final state and make use of the latest $\sqrt{s} = 13$ TeV CMS search for a narrow di-Higgs resonance in this channel  
with $35.9$ fb$^{-1}$ of integrated luminosity~\cite{Sirunyan:2018zkk}.
In this section, we validate our implementation of the CMS analysis by fixing $m_{H_2}=125\;$GeV and reproduce both the signal selection efficiencies and the 95\% C.L. cross-section upper limits reported in~\cite{Sirunyan:2018zkk}, before extending our analysis to the 2D mass plane ($m_{H_1}$, $m_{H_2}$) in section~\ref{sec:enlarging}. 

\subsection{Validation of the selection efficiencies for the signal}

The CMS collaboration reports 
the signal efficiencies at various stages of the spin-$0$ analysis event selection, 
namely from trigger level up to the signal region (SR) definition. The search defines two kinematic regions that feature different event selection criteria: a low-mass-region (LMR) for masses $m_{H_1} \in [250,\, 650]$ GeV, and a medium-mass-region (MMR) for masses $m_{H_1} \in [550,\, 1200]$ GeV\footnote{For $m_{H_1} > 1200$ GeV, the angular separation between the two $b$-quarks from a Higgs decay is typically too small to satisfy jet isolation criteria, causing a large drop in the signal selection efficiencies. A different analysis strategy making use of jet-substructure techniques is needed in this regime.}. The transition region $m_{H_1} \sim 580$ GeV is determined by the respective sensitivities of the LMR and MMR selection strategies~\cite{Sirunyan:2018zkk}.

Events are selected with an online trigger that requires either of the following conditions to be satisfied
\begin{itemize}
\item[{\it i)}] Four reconstructed jets of $p_T > 30$ GeV and $|\eta|<2.4$
of which two satisfy $p_T>90\;$GeV and at least three $b$-tagged jets.
\item[{\it ii)}] Four reconstructed jets of $p_T > 45$ GeV of which at least three are 
$b$-tagged. 
\end{itemize}

The analysis then requires all four selected jets to be $b$-tagged\footnote{The {\tt DeepCSV} $b$-tagging medium working point used yields an average $b$-tagging efficiency of 68\% and respective mistag probabilities for $c$-jets and light-jets of 12\% and 1.1\%~\cite{Sirunyan:2017ezt} (see Appendix~\ref{app:btagging} for details).} and be within $\left|\eta \right| < 2.4$. This initial selection stage is labelled {\sl 4$b$} and is common to both the LMR and MMR categories.
For the LMR selection the analysis then identifies two 125 GeV Higgs boson candidates by pairing the $b$-jets and requiring $|m_{b\bar{b}}-120\;{\rm GeV}| < 40$~GeV for each pair, while for the MMR selection the two $b$-jet pairs must satisfy $\Delta R_{bb} < 1.5$. 
This selection is named ``$HH$ candidate"\footnote{For both LMR and MMR selection categories, in case of multiple $HH$ candidate combinations in an event, the combination that minimizes $\chi$ as defined in~Eq.~\eqref{chi_HH} is chosen. We also note that 
$\Delta R_{bb}$ depends only on the mass ratio $m_{H_1}/m_{h_{\mathrm{SM}}}$~\cite{Gouzevitch:2013qca}, and as such the ratio of signal efficiencies at {\sl 4$b$} and $HH$ candidate stage could for the MMR category in principle be approximately extrapolated to a 2D mass plane, modulo acceptance effects that depend on the individual scalar masses.}. 
Finally, the SR is defined in the two dimensional space of the reconstructed masses of the lighter Higgs boson candidates, $m_{H_2^1}$ and $m_{H_2^2}$, as the circular region with $\chi < 1$, where $\chi$ is defined as
\begin{equation}
\label{chi_HH}
\chi = \sqrt{\left(\frac{m_{H_2^1} - C}{R} \right)^2 + \left(\frac{m_{H_2^2} - C}{R} \right)^2}.
\end{equation}
The values of the parameters $C$ and $R$ are set to $(C,\, R) = (120,\,20)\;$GeV and $(C,\, R) = (125,\,20)\;$GeV for the LMR and MMR category respectively. 
We note that further improvements in the Higgs boson mass resolution through multivariate regression techniques applied by the CMS analysis are not included in our 
analysis. These increase the sensitivity of the CMS analysis by 5 -- 20\% depending on the mass hypothesis~\cite{Sirunyan:2018zkk}, and thus our validation is expected to yield a potential mismatch of at least that order.

For our validation we have implemented the relevant interactions for the spin-0 BSM state in the {\tt Feynrules} package~\cite{Alloul:2013bka} and 
generated hard-scattering events through the  {\tt Madgraph5$\_$aMC@NLO} platform~\cite{Alwall:2014hca}. These events have been generated at leading order (LO) with fixed widths of 10 GeV and 1 GeV for $H_1$ and $H_2$ respectively and up to two additional jets in the matrix element. The matching and merging between hard-scattering and parton shower has been performed via the {\tt MLM} procedure\footnote{We have set {\tt xqcut}={\tt qcut}=$m_{H_1}/4$ GeV.}~\cite{Mangano:2006rw} with {\tt PYTHIA8}~\cite{Sjostrand:2007gs} using the shower-$k_T$ scheme. Finally, {\tt Delphes}~\cite{deFavereau:2013fsa} is used for a simulation of the CMS detector performance which also makes use of the {\tt Fastjet}~\cite{Cacciari:2011ma} algorithm to cluster {\tt anti-$k_T$}~\cite{Cacciari:2008gp} jets with radius $R = 0.4$. 
A crucial ingredient in this last step concerns the 13 TeV CMS $b$-tagging efficiencies, as well as the $c$-jet and light-jet mis-tag rates which are functions of the jet $p_T$ and $\eta$. We have modeled these rates using the information from~\cite{Sirunyan:2017ezt},
assuming the performance of the {\tt DeepCSV} $b$-tagging algorithm for the same operating point as used in~\cite{Sirunyan:2018zkk} (see Appendix~\ref{app:btagging} for details).

\begin{figure}[t]
\begin{center}
\includegraphics[width=0.64\textwidth]{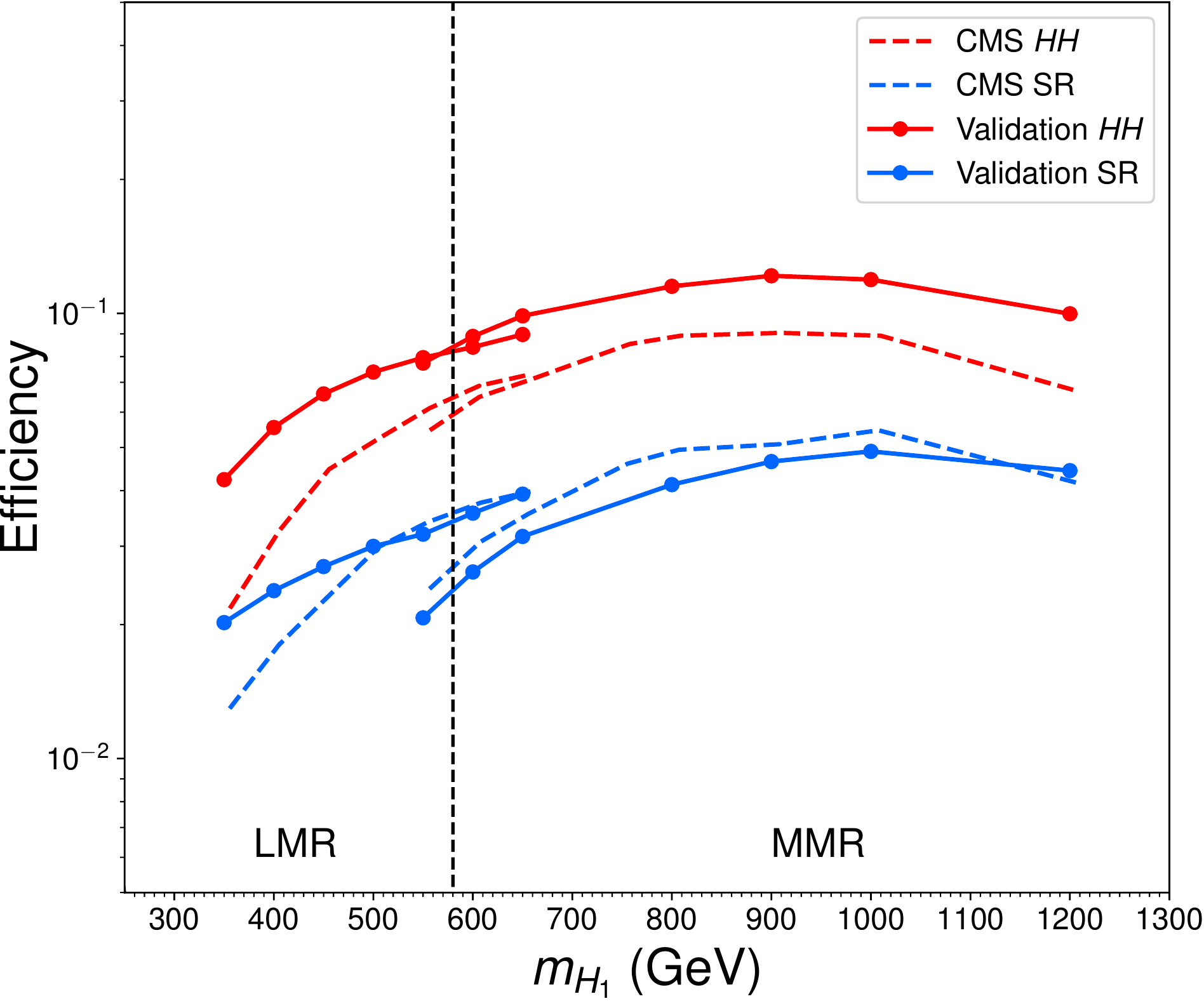}\hfill
\caption{\small Signal efficiencies as a function of $m_{H_1}$: 
 {\sl $HH$} (red) and {\sl $SR$} (blue) stages for spin-0 resonant di-Higgs production in 
the $b \bar{b} b \bar{b}$ final state. Solid lines correspond to our simulation, while dashed lines correspond to the efficiencies from the 
13 TeV CMS analysis~\cite{Sirunyan:2018zkk}. The vertical dashed black line at $m_{H_1} = 580$ GeV marks the boundary between the LMR and MMR analysis categories.}
\label{Fig_HAA_effs}
\end{center}

\vspace{-3mm}

\end{figure}

Our simulated signal efficiencies at the $HH$ and SR stages are shown in Fig.~\ref{Fig_HAA_effs} for both the LMR and the MMR regions, together with the corresponding CMS efficiencies from~\cite{Sirunyan:2018zkk}. Overall, we find the agreement between our validation efficiencies and those reported by the $\sqrt{s} = 13$ TeV $p p \to H_1 \to h_{\rm SM} h_{\rm SM} \to b \bar{b} b \bar{b}$ CMS analysis to be better than 50\% (except for the very low LMR masses, where the agreement is worse and the mismatch at the $H H$ stage can reach 90\%). The mostly moderate mismatch at the {\sl $HH$} selection level can be understood from our use of a fast detector simulation and our relatively limited information in the modeling of $b$-tagging efficiencies, as discussed in Appendix~\ref{app:btagging}. The agreement is nevertheless very good (better than 15\%) for the SR selection with $m_{H_1} > 450$~GeV, {\emph{i.e.}} for the whole mass region of the MMR category and part of the LMR category\footnote{This may be a result of our slight overestimate of $H H$ efficiencies providing a partial compensating effect to the sensitivity improvement from the use of regression techniques in the SR by the CMS analysis.}, as can be seen from Fig.~\ref{Fig_HAA_effs}.

\subsection{Validation of the cross section upper limits}

\begin{figure}[h]
\begin{center}
\includegraphics[width=0.49\textwidth]{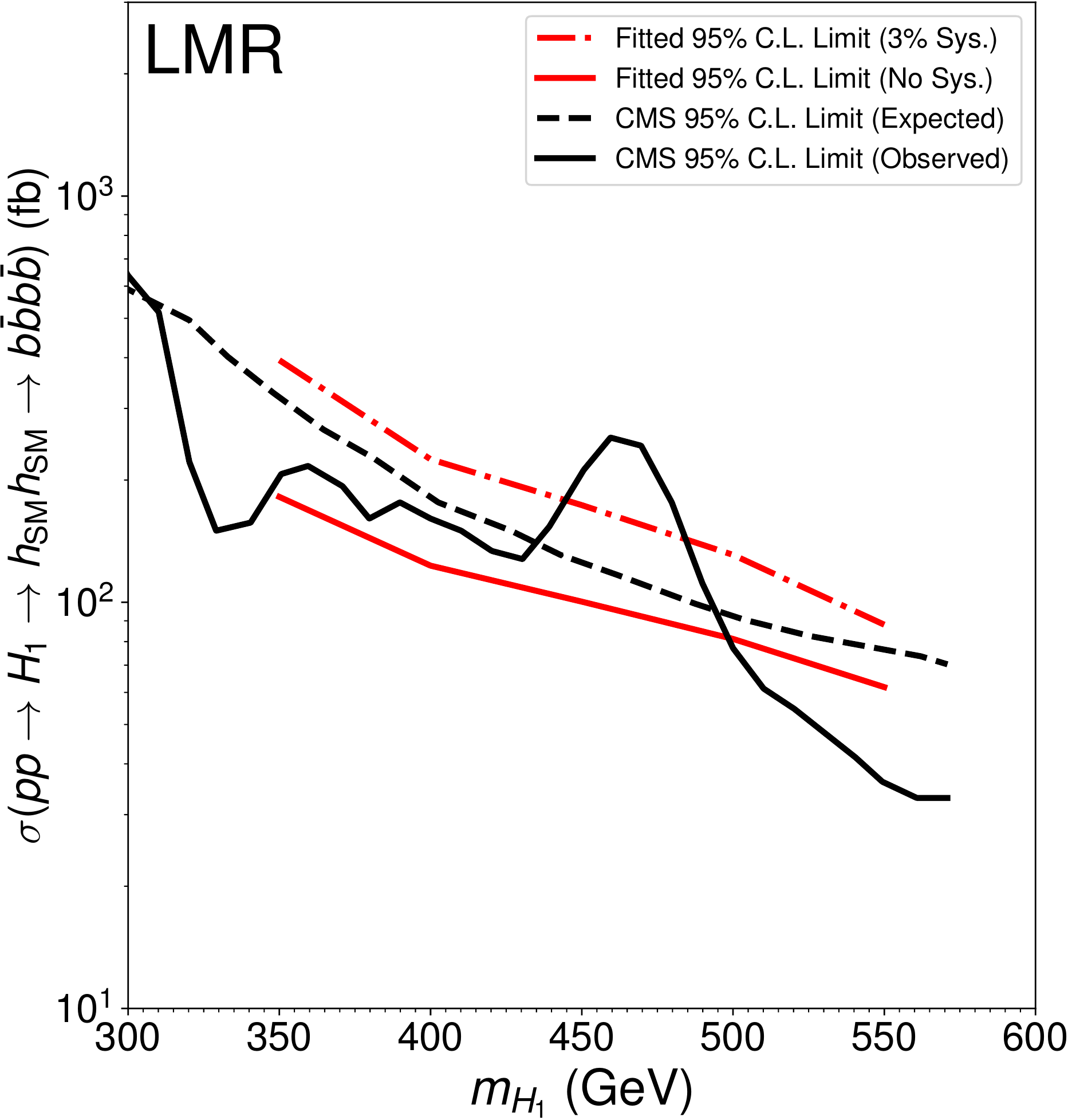}\hfill
\includegraphics[width=0.49\textwidth]{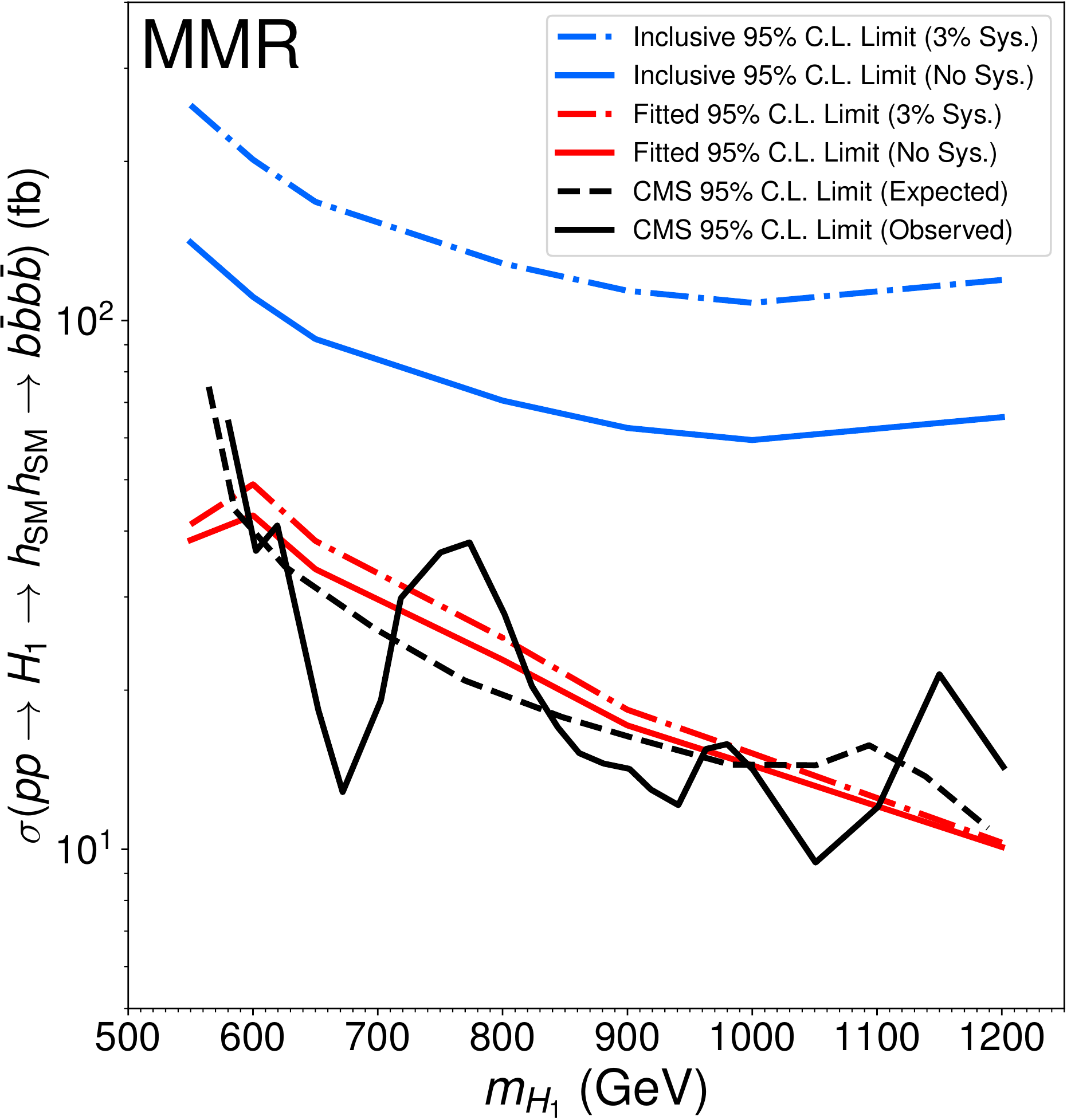}\hfill


\caption{\small 95\% C.L.
upper limits on the $p p \to H_1 \to h_{\rm SM} h_{\rm SM} \to b \bar{b} b \bar{b}$ cross section (in fb) in the LMR (left) and MMR (right) regions. Black lines correspond to the observed (solid) and expected (dashed) limits 
from the CMS analysis~\cite{Sirunyan:2018zkk}. The solid (dash-dot) blue line, only available for MMR, is the {\sl inclusive} limit considering the total event yield in the SR defined by Eq.~\eqref{chi_HH}, assuming no systematic uncertainty (3\% systematic uncertainty) on the SM background. The solid (dash-dot) red line correspond to the {\sl fitted} limit by considering only SM background events in the SR and within a window of $\pm 2\, \Gamma_{H_1}$ around the considered signal resonance mass $m_{H_1}$, assuming no systematic uncertainty (3\% systematic uncertainty) on the SM background (see main text for details).}
\label{Fig_HAA_UL}
\end{center}

\vspace{-3mm}

\end{figure}

Besides reproducing the selection efficiencies of the CMS experimental analysis, it is crucial to check that our procedure can provide upper limits on the signal cross section consistent with those obtained by the CMS collaboration. CMS provides the inclusive background yield, dominated by QCD multi-jet processes, at the SR selection stage in the ($m_{H_{2}^1}$, $m_{H_{2}^2}$) plane (recall Eq.~\eqref{chi_HH}) for the MMR category~\cite{Sirunyan:2018zkk}, while such information is not available for the LMR category. This background yield is independent of the value of $m_{H_1}$ considered and in the region defined by Eq.~\eqref{chi_HH}, approximately 2630 SM background events are found. This results in a 95\% C.L. upper limit on the signal event yield of $\sim 105$ events considering only the statistical uncertainty on the SM 
background, using a significance measure of $N_S/\sqrt{N_S+N_B}$ (with $N_S$ and $N_B$ respectively the number of signal and SM background events).
The derived limits on the inclusive $p p \to H_1 \to h_{\rm SM} h_{\rm SM} \to b \bar{b} b \bar{b}$ signal cross section are shown in the right panel of Fig.~\ref{Fig_HAA_UL} (solid blue line). 
In the presence of SM background systematic uncertainties, our significance measure gets modified to $N_S/\sqrt{N_S+N_B + u_{B}^2 N_B^2}$, with $u_B$ the SM background systematic error, and
we also derive the corresponding inclusive limits assuming a $3\%$ systematic uncertainty on the SM background, illustrated in the right panel of Fig.~\ref{Fig_HAA_UL} as a dot-dashed blue line.
The 3\% value chosen for the background systematics is mildly conservative for the MMR category and clearly shows the degrading of the limits due to systematic uncertainties in Fig.~\ref{Fig_HAA_UL}.
This value is chosen based on a comparison of our analysis with HE-LHC projections for resonant di-Higgs production~\cite{CMS:2017cwx}, where we find that a value of 2\% reproduces the projected limits quoted therein (see section~\ref{sec:HL_LHC_2} for more details).

The above inclusive limits for the MMR category (we note again that for the LMR category it is not possible to extract an inclusive limit from public CMS data) are a factor $\sim 3 - 4$ weaker than the CMS 95\% C.L. upper limit on the signal cross section from~\cite{Sirunyan:2018zkk}, shown in the right panel of Fig.~\ref{Fig_HAA_UL} as a solid (dashed) black line for the 95\% C.L. observed (expected) limit. 
The reason is that the limit 
is not computed in an inclusive manner (that is, solely from the signal and SM background event yield after SR selection defined by Eq.~\eqref{chi_HH});
rather, it is extracted by fitting the SM QCD background distribution after the SR selection as a function of the invariant mass of the four $b$-jet system $m_{4b}$, and considering only SM background events within a certain width around the signal hypothesis $m_{4b} \sim m_{H_1}$. Using the SM background $m_{4b}$ distribution after SR selection provided by the CMS analysis~\cite{Sirunyan:2018zkk} for the MMR (provided in~\cite{Sirunyan:2018zkk} for $m_{H_1} > 550$ GeV) and LMR categories, and defining a $\pm 2\, \Gamma_{H_1}$ mass window\footnote{Specifically, we adopt $[m_{H_1},\Gamma_{H_1}]=[450,\;12.3],\;[710,\;21.3],\;[915,\;31.4]$ GeV, as considered in~\cite{Sirunyan:2018zkk} and interpolate linearly between them.} around each $m_{H_1}$ signal hypothesis, 
we obtain the corresponding {\sl fitted} 95\% C.L. upper limits on the 
$p p \to H_1 \to h_{\rm SM} h_{\rm SM} \to b \bar{b} b \bar{b}$ signal cross section, both without systematic uncertainties and again assuming a 3\% systematic uncertainty on the SM background. These are respectively shown in Fig.~\ref{Fig_HAA_UL} for the MMR (right panel) and LMR (left panel) categories as solid red lines (no systematics) and dash-dot red lines (3\% systematics), showing good agreement with the expected 95\% C.L. upper limits reported by the CMS analysis. These results validate our extrapolation of the CMS analysis~\cite{Sirunyan:2018zkk} to search for BSM scalars, which we do in the next section.

\section{Searching for new scalars via $p p \to H_1 \to H_2 H_2 \to b \bar{b} b \bar{b}$}
\label{sec:enlarging}

Having validated our implementation of the CMS experimental analysis, we can now proceed to extend the search to the ($m_{H_1}$, $m_{H_2}$) mass plane. For the event generation we follow the procedure discussed in the previous section, within a 2D mass grid defined as follows:

\begin{itemize}

\item $m_{H_1}$ is varied in the range $[300,\, 1000]$~GeV in steps of 50 GeV.
    
\item For each $m_{H_1}$ value, $m_{H_2}$ is varied in the range $[65\;{\rm GeV},\, m_{H_1}/2]$, taking ten equally spaced values.      
    
\end{itemize}

The various parameters defining the SR selection in Eq.~\eqref{chi_HH} for LMR and MMR categories need to be modified accordingly. For the MMR category, we maintain $R = 20\;$GeV and set $C = m_{H_2}$, while for the LMR category we also keep $R = 20$ GeV and set instead $C = (120/125) \times m_{H_2}$. In addition, for the LMR the $HH $ candidate selection on the $b$-jet pairs needs to be modified to $|m_{b\bar{b}} - (120/125)\, m_{H_2}| < \mathrm{Max}[20\,\mathrm{GeV},\, (40/125)\, m_{H_2}]$. These modifications match the CMS analysis $H H$ and SR selection criteria for $m_{H_2} = 125\;$GeV, while prodiving a natural generalization of those for other values of $m_{H_2}$. 

\vspace{1mm}

For the estimate of the SM QCD multi-jet background, we first extrapolate the SM background event yield in the SR as a function of $m_{H_2}$ to the region $m_{H_2} > 300$ GeV using a smoothly falling exponential fit\footnote{We observe that such a fit provides a very good description of the measured SM background yield in the SR for $125\,\,\mathrm{GeV}< m_{H_2} < 300\,\,\mathrm{GeV}$.}. Then, we adopt the following procedure: 
\begin{itemize}

\item In the MMR category with $m_{H_1} > 550$~GeV, the {\sl fitted} SM background event yield is computed as described in the previous section, with an additional overall rescaling of the (MMR) SM background $m_{4b}$ distribution. The factor is determined by the ratio of the inclusive background yield in the SR with $C = m_{H_2}$ over the SR background yield for $C = 125$ GeV. This procedure assumes that the SM background $m_{4b}$ shape remains approximately unchanged and only its overall normalization varies when the SR selection from Eq.~\eqref{chi_HH} is redefined by setting $C = m_{H_2}$.

\item In the LMR category with $m_{H_1} < 550$~GeV\footnote{In the rest of the paper, we consider the boundary between LMR and MMR categories at $m_{H_1} = 550$~GeV.}, we follow the same strategy as for the MMR category above, performing the aforementioned overall rescaling of the SM background $m_{4b}$ distribution (now for the LMR category).  However, since no inclusive SM background yield after SR selection is provided by the CMS analysis for the LMR category, we use the same rescaling factor (as a function of $m_{H_2}$) as for the MMR category.

\item For the MMR category with $m_{H_1} < 550$~GeV, it is not possible to apply the above strategy, since the SM background $m_{4b}$ distribution for the MMR category is not provided by the CMS analysis~\cite{Sirunyan:2018zkk} in this region (and thus the {\sl fitted} 95\% C.L. upper limits on the signal cross section cannot be derived). Instead, we use the total SM QCD background yield in the SR (redefined by $C = m_{H_2}$) to obtain the inclusive 95\% C.L. upper limits on the signal cross section. 

\end{itemize}

These procedures for the estimate of the SM background are nevertheless expected to fail both for $m_{H_1} < 300$ GeV and $m_{H_1} \to 2 m_{H_2}$, since in these regions the CMS measured multi-jet data do not follow a smoothly falling distribution, but rather display a kinematic feature near the threshold region, driven by the kinematic selection of the analysis (mainly trigger effects)~\cite{Sirunyan:2018zkk}. This  leads to a feature that depends on $m_{H_2}$, and peaks around 300 GeV in the CMS analysis. We avoid being near these regions of the $(m_{H_1}, m_{H_2})$ plane by imposing $m_{H_1} > 300$ GeV and  $m_{H_1} > 2 m_{H_2} + 25$ GeV in our analysis. 

\begin{figure}[ht!]
\centering

$\vcenter{\hbox{\includegraphics[width=0.74\textwidth]{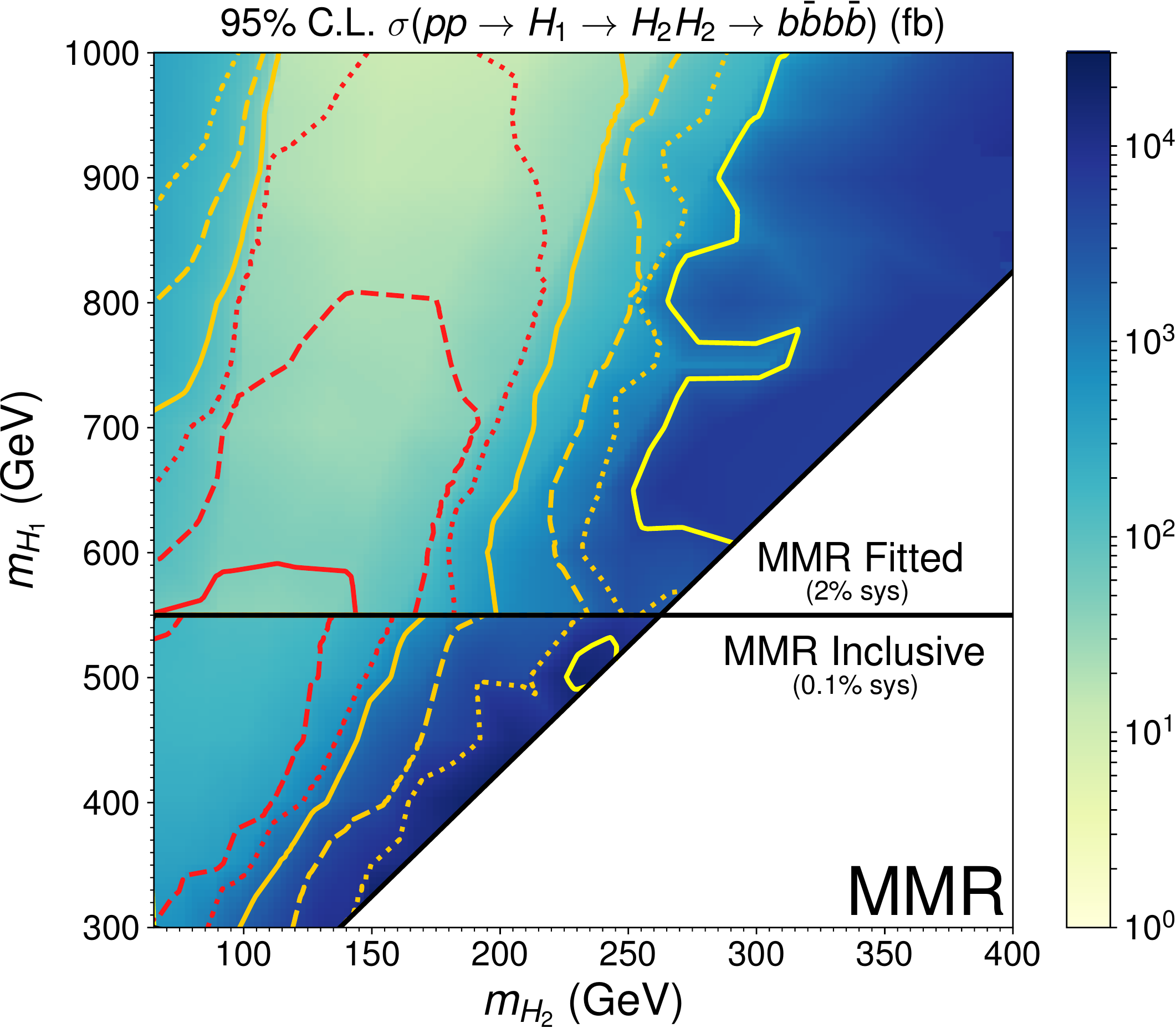}}}$
\hspace{4mm}
$\vcenter{\hbox{\includegraphics[width=0.17\textwidth]{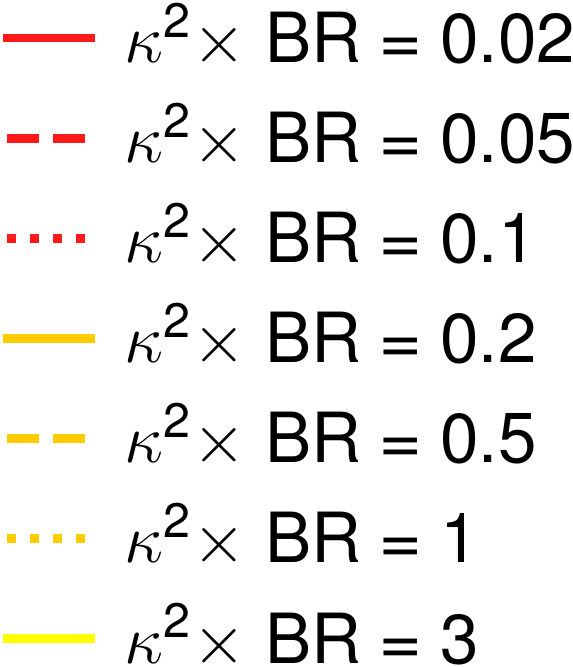}}}$

\caption{\small 95\% C.L. upper limit on the $pp\to H_1 \to H_2 H_2 \to b\bar b b \bar b$ signal cross section (in fb) in the ($m_{H_2}, \, m_{H_1}$) 
plane for the MMR category, for $m_{H_1} > 550$ GeV ({\sl fitted} limit, assuming a $2\%$ systematic uncertainty) and for $m_{H_1} < 550$ GeV (inclusive limit, assuming a $0.1\%$ systematic uncertainty), extending the CMS analysis~\cite{Sirunyan:2018zkk}. The various contours correspond to the 95\% C.L. upper limits on the 
sensitivity $\kappa^2 \times {\rm BR}$ from Eq.~\eqref{eq:kappaBR}, see text for details.}
\label{fig:MMR_extension}
\end{figure}

\begin{figure}[t!]
\centering

$\vcenter{\hbox{\includegraphics[width=0.73\textwidth]{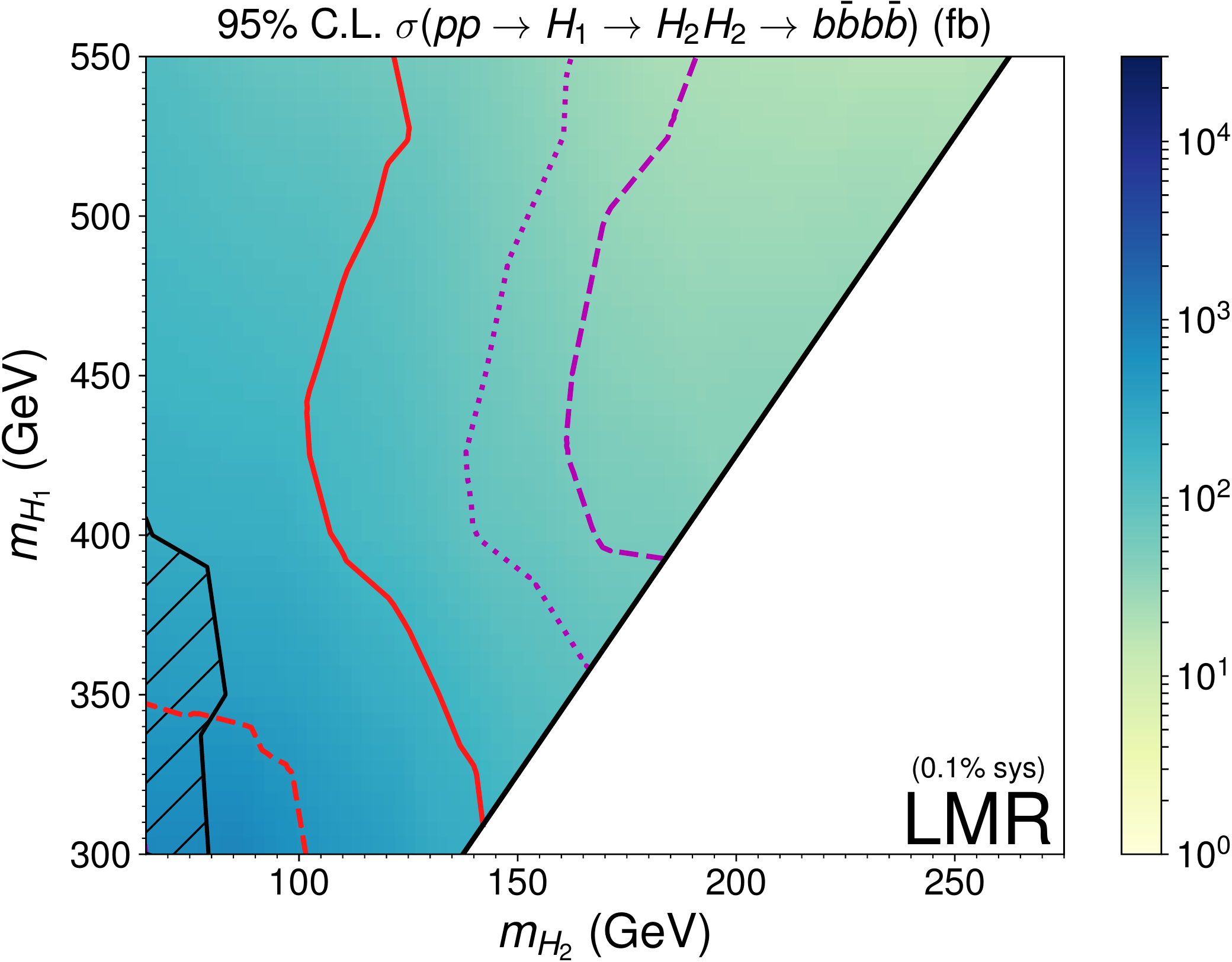}}}$
\hspace{4mm}
$\vcenter{\hbox{\includegraphics[width=0.19\textwidth]{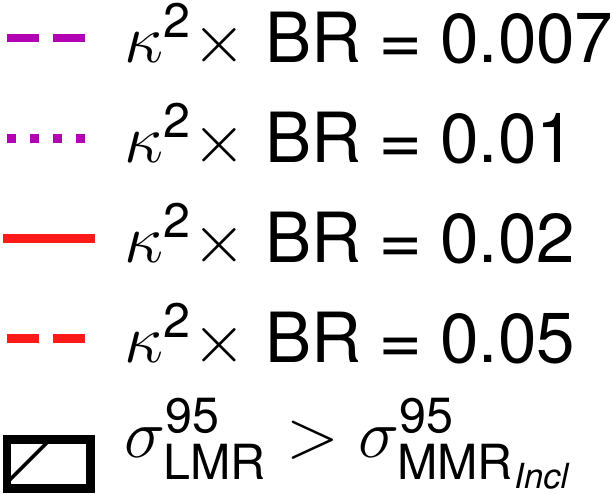}}}$

\caption{\small ({\sl fitted}) 95\% C.L. upper limit on the $pp\to H_1 \to H_2 H_2 \to b\bar b b \bar b$ signal cross section (in fb) 
in the ($m_{H_2}, \, m_{H_1}$) plane for the LMR category ($m_{H_1} < 550$ GeV, assuming a $0.1\%$ background systematic uncertainty), extending the CMS analysis~\cite{Sirunyan:2018zkk}. In the hatched region, the MMR inclusive 
limits from Fig.~\ref{fig:MMR_extension} are stronger than the LMR limits. The various contours correspond to the 95\% C.L. upper limits on the 
sensitivity $\kappa^2 \times {\rm BR}$ from Eq.~\eqref{eq:kappaBR}.}
\label{fig:LMR_extension}

\vspace{-2mm}
\end{figure}

The 95\% C.L. upper limits on the $pp\to H_1 \to H_2 H_2 \to b\bar b b \bar b$ cross section are shown in Fig.~\ref{fig:MMR_extension} for the MMR category 
and in Fig.~\ref{fig:LMR_extension} for the LMR category as color coded heat maps.
For the MMR category, Fig.~\ref{fig:MMR_extension} shows both the {\sl fitted} limit for $m_{H_1} > 550$ GeV assuming a $2\%$ SM background systematic uncertainty and the inclusive limit for $m_{H_1} < 550$ GeV with a SM background $0.1\%$ systematic uncertainty. The same $0.1\%$ background systematic uncertainty is assumed for the LMR category. 
These choices for the SM background systematic uncertainty are motivated in section~\ref{sec:HL_LHC_2}, and are conservative given the information from the CMS analysis~\cite{Sirunyan:2018zkk}.
The SM background systematic uncertainties are driven by the background modeling, resulting in a smaller error at lower invariant masses given the larger statistics in that region, which explains the difference in systematics between MMR and LMR categories.  
We also show the 95\% C.L. signal cross section upper limits with no SM background systematic uncertainties in Appendix~\ref{app:H1H2H2nosys}.
By comparing the results from Figs.~\ref{fig:MMR_extension} and~\ref{fig:LMR_extension} with Fig.~\ref{fig:AppendixB_Limits} (top-left panel) in Appendix~\ref{app:H1H2H2nosys}, we see that the effect of background systematics is not very important for current signal upper limits, which are at present statistically dominated.  

Comparing the inclusive signal 95\% C.L. upper limits for the MMR category ($m_{H_1} < 550$ GeV) to those for the LMR category from Fig.~\ref{fig:LMR_extension}, we see the latter are much stronger except for the small region $m_{H_2} \lesssim 80$ GeV, $m_{H_1} \lesssim 400$ GeV (see Fig.~\ref{fig:LMR_extension}). 
We thus omit from now on the use of the MMR inclusive results for $m_{H_1} < 550$~GeV and use the LMR and MMR fitted 95\% C.L. signal cross section upper limits respectively for $m_{H_1}$ lighter and heavier than $550$~GeV. 

We can then parametrize our BSM cross section as
\begin{equation}
\label{eq:kappaBR}
\sigma(pp\to H_1 \to H_2 H_2 \to b\bar b b \bar b) = \hat{\sigma}_{H_1} \times \kappa^2 \times {\rm BR}
\end{equation}
with $\hat{\sigma}_{H_1}$ the inclusive production cross section of a SM-like Higgs boson with mass $m_{H_1}$, $\kappa^2$ an effective rescaling factor with respect to the SM-like Higgs boson cross section and BR $=$ BR$(H_1 \to H_2 H_2 \to b\bar b b \bar b)$. Through this parametrization we can translate the derived upper limits on the cross sections into limits on $\kappa^2 \times  {\rm BR}(H_1 \to H_2 H_2 \to b\bar b b \bar b)$. These limits, shown as isocontours in Figs.~\ref{fig:MMR_extension} and~\ref{fig:LMR_extension}, serve as a reference point for understanding the possible impact of our search in specific BSM models, which are discussed in detail in section~\ref{sec:models}.

From Fig.~\ref{fig:LMR_extension} we see that in the LMR category, for a fixed value of $m_{H_1}$ the sensitivity of the search increases 
with the mass $m_{H_2}$. On the other hand, for the MMR category and a fixed $m_{H_1}$, increasing $m_{H_2}$ from 65 GeV results in an 
increase in sensitivity up to an optimal value of $m_{H_2} \sim 120 - 180$ GeV (depending on the value of $m_{H_1}$), above which the sensitivity 
of the search drops and the {\sl fitted} 95\% C.L. limit on the signal cross section quickly becomes very large, as shown in Fig.~\ref{fig:MMR_extension}. 
This behaviour of the sensitivity for the LMR and MMR categories can be understood from the interplay between the signal acceptance 
and the SM multi-jet background yield in the SR. The SM background yield in the SR decreases rapidly as $m_{H_2}$ increases, which explains the 
behaviour observed for the LMR category, as well as the initial growth in sensitivity for $m_{H_2} > 65$ GeV in the MMR category. For the MMR category, the SR acceptance decreases for the BSM signal as $m_{H_2}$ increases for a fixed $m_{H_1}$ and eventually overcomes the decrease in the SR background yield, and the sensitivity drops again. This however does not occur for the LMR category, which retains sensitivity to the $2\,m_{H_2} \to m_{H_1}$ region.

\vspace{2mm}

Overall, it is interesting to note that values of $\sigma(pp\to H_1 \to H_2 H_2 \to b\bar b b \bar b)$ in the ballpark of 
several tens of fb can be accessed for the MMR category in the optimal region $m_{H_2} \sim 100 - 150$ GeV, yielding a sensitivity to 
$\kappa^2 \times \mathrm{BR} \lesssim 0.02 - 0.1$ depending on the value of $m_{H_1}$. At the same time, the sensitivity to $\kappa^2 \times \mathrm{BR}$ for the LMR category reaches values as low as $7\times 10^{-3}$, with most of the LMR parameter space being constrained to $\kappa^2 \times \mathrm{BR} < 0.05$ at 95\% C.L. 
This suggests that our proposed search may indeed be sensitive to new scalars in realistic BSM scenarios. We discuss this in more detail in section~\ref{sec:models}.

\section{$p p \to H_1 \to H_2 H_2 \to b \bar{b} b \bar{b}$ at the HL-LHC and future hadron colliders}
\label{sec:enlarging_2}

In this section we discuss the extrapolation of the current limits derived in section~\ref{sec:enlarging} to several future proton-proton machines, differing in their center-of-mass energy $\sqrt{s}$ and in the targeted total integrated luminosity. The different future setups that we consider\footnote{We also note that, as of now, ATLAS and CMS have recorded $\sim 150\;$fb$^{-1}$ each, currently under analysis.} are shown in Table~\ref{tab:futcoll}. We furthermore study the impact of systematic uncertainties, and briefly comment on the role of trigger and $b$-tagging efficiencies, which are critical for this final state.

\begin{table}[h!]
\begin{center}
\begin{tabular}{|l|c|c|c|}
\hline
  & \multicolumn{1}{l|}{$\sqrt{s}$ {[}TeV{]}} & $\int {\cal L}$ {[}fb$^{-1}${]} & Reference                                 \\ \hline
HL-LHC & 14                                        & $3\times 10^3$                            &   \cite{CMS:2017cwx}                                       \\
HE-LHC & 27                                        & $15\times 10^3$                           & \cite{Abada:2019ono}    \\
FCC-hh & 100                                       & $30\times 10^3$                           & \cite{Benedikt:2018csr} \\ \hline
\end{tabular}
\caption{Future hadron collider scenarios considered in the text.} 
\label{tab:futcoll}
\end{center}

\vspace{-6mm}

\end{table}

\subsection{General procedure for the extrapolation to higher $\sqrt{s}$}
\label{sec:HL_LHC_1}

\begin{figure}[t]
\begin{center}
\includegraphics[width=0.67\textwidth]{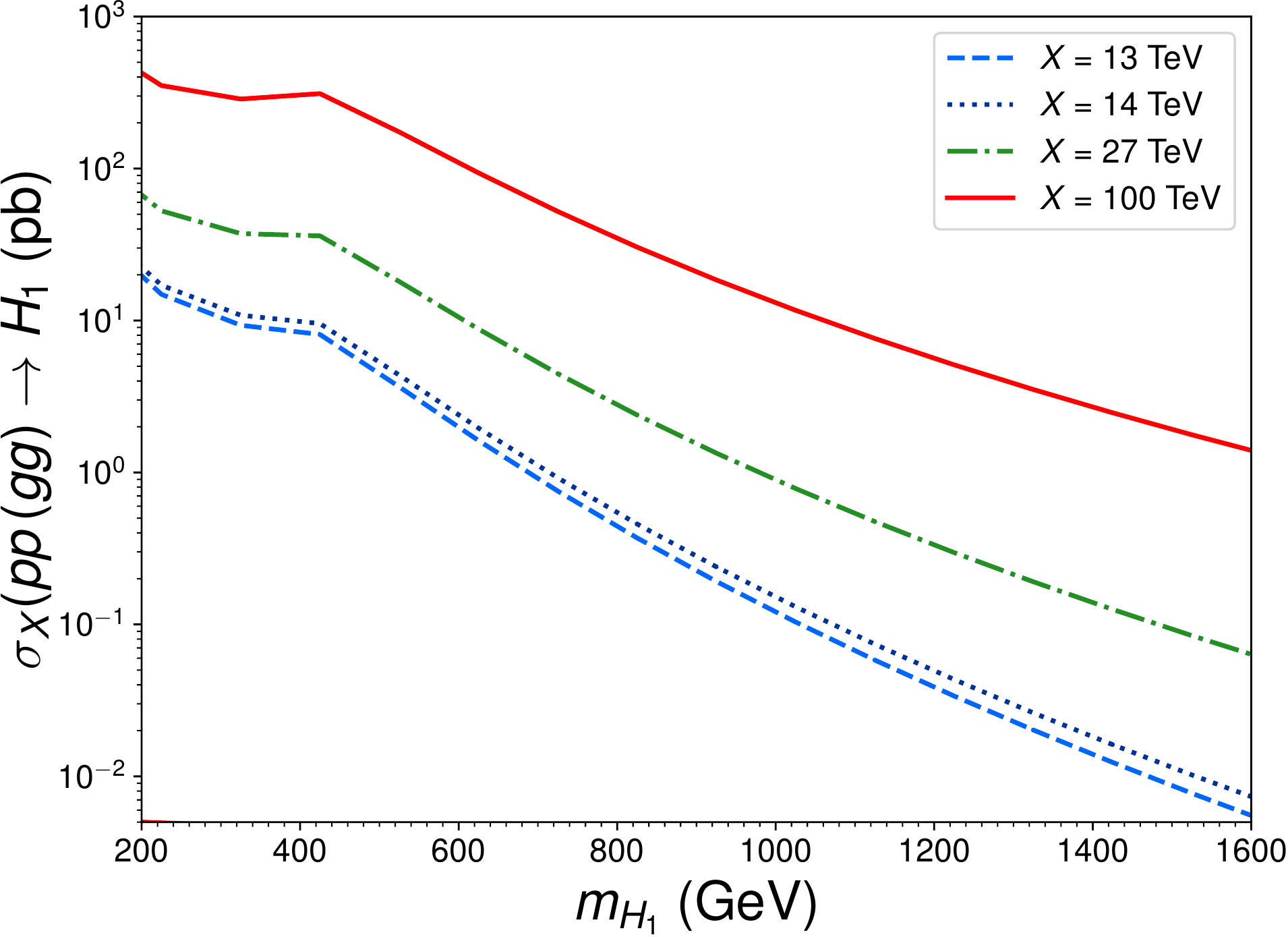}\hfill
\caption{\small Gluon-fusion production cross section of a SM-like Higgs $H_1$ as a function of its mass $m_{H_1}$, for hadron colliders with different center of mass energies $\sqrt{s} = X$: 13 TeV (light-blue, dashed), 14 TeV (dark-blue, dotted), 27 TeV (green, dot-dashed) and 100 TeV (red, solid).}
\label{fig:XS_colliders}
\end{center}

\vspace{-2mm}

\end{figure}

The main challenge in computing the reach for hadron colliders with higher centre of mass energy  is the scaling of the multi-jet background. This is estimated in a data-driven way by the CMS and ATLAS experimental collaborations and cannot be reliably simulated within our framework. Both the SM background and the signal increase for higher collider energies, while the parton shower is expected to provide a larger number of significantly harder jets, thus possibly changing the kinematic features of the events.
Here we perform our analysis by naively assuming that both the signal and the SM background cross sections scale roughly by the same amount, given by the ratio of inclusive $gg \to H_1$ production cross sections at $\sqrt{s} = 13$ TeV and $\sqrt{s} = X$ TeV, namely
\be\label{eq:rescaling}
r_{X}= \frac{\sigma_{X} (gg \to H_1)}{\sigma_{13} (gg \to H_1)}\, \, .
\ee
This assumption is motivated by the fact that the 
overall partonic centre of mass energy of both the signal and the SM background after the full event selection will be peaked around $\sqrt{\hat s} \sim m_{H_1}$ and is valid to the extent that the $gg$ luminosity dominates the multijet rate. 
To compute the rescaling factor $r_X$ we use 
\texttt{SusHi v1.7.0}~\cite{Harlander:2012pb,Harlander:2016hcx}, which gives NNLO accuracy for the production of a SM-like Higgs boson in the infinite top mass limit~\cite{Harlander:2002wh,Harlander:2003ai,Actis:2008ug,Harlander:2005rq,Chetyrkin:2000yt}. These cross sections are reported in Fig.~\ref{fig:XS_colliders} for the various collider scenarios of Table~\ref{tab:futcoll}. 
We further assume that the acceptance and selection efficiencies of the proposed search remain approximately constant for the different collider scenarios considered. This is expected to be a good approximation for the case of the HL-LHC (modulo improvements in trigger and $b$-tagging efficiencies, which we discuss at the end of section~\ref{sec:HL_LHC_2}), while it will not be very accurate for the HE-LHC and FCC-hh.
In any case, our results should be interpreted as a conservative first estimate of the sensitivity of the $p p \to H_1 \to H_2 H_2 \to b\bar b b \bar b$ search channel at the HL-LHC and future hadron colliders, bearing in mind that future analyses may improve upon this estimate.

In general, the expected improvement in sensitivity on the $\kappa^2\times{\rm BR}$ factor of Eq.~\eqref{eq:kappaBR} for a future collider with respect to the current $\sqrt{s}=13\;$TeV limits from section~\ref{sec:enlarging} may be simply written as
\be
\label{eq:sensitivity_resc}
I^{-1} \equiv \frac{1}{r_X} \times \frac{\sigma_{X}^{95\%\mathrm{C.L.}}}{\sigma_{13}^{95\%\mathrm{C.L.}}} 
= \frac{\left. \kappa^2 \times \mathrm{BR}\right|_X^{ 95\% \mathrm{C.L.}}}{\left. \kappa^2 \times \mathrm{BR}\right|_{13}^{ 95\% \mathrm{C.L.}}}
\ee
with $\sigma_{X}^{95\%{\rm{C.L.}}}/\sigma_{13}^{95\%{\rm{C.L.}}}$ the ratio of 95\% C.L. signal cross section upper limits at $\sqrt{s}=13\;$TeV and $\sqrt{s}=X$ TeV, and $I > 1$ yielding an improvement in sensitivity. 
Under the assumptions made in this section and for $N_B \gg N_S$ the ratio 
$\sigma_{X}^{95\%{\rm{C.L.}}}/\sigma_{13}^{95\%{\rm{C.L.}}}$ is given in the 
absence of systematic uncertainties simply by
\be
\label{eq:ul_resc}
\frac{\sigma_{X}^{95\%{\rm{C.L.}}}}{\sigma_{13}^{95\%{\rm{C.L.}}}}=\sqrt{r_X\,\frac{{\cal L}_{13}}{{\cal L}_{X}}}\,\,.
\ee
As an example, for $m_{H_1} = 1$ TeV the ratio $\sigma_{X}^{95\%{\rm{C.L.}}}/\sigma_{13}^{95\%{\rm{C.L.}}}$ is given, in the absence of systematic uncertainties, by $0.12$, $0.13$ and $0.36$ respectively for HL-LHC, HE-LHC and FCC-hh, which would then result sensitivity improvements of in respective $I = 10$, $50$ and $300$ respectively.
Nevertheless, the impact of systematic uncertainties on the ratio $\sigma_{X}^{95\%{\rm{C.L.}}}/\sigma_{13}^{95\%{\rm{C.L.}}}$ may be important, and we discuss this in more detail in section~\ref{sec:HL_LHC_2}.

\subsection{Validation \& Extrapolation to HL-LHC}
\label{sec:HL_LHC_2}

Before presenting our results for the HL-LHC extrapolations, we validate our procedure by comparing our findings for the case $m_{H_2}=125\;$GeV with the official CMS projections for the HL-LHC in the $h_{\mathrm{SM}} h_{\mathrm{SM}} \to b \bar{b} b \bar{b}$ channel~\cite{CMS:2017cwx}. The CMS collaboration reports projected 95\% C.L. cross section upper limits of 
46, 7.3 and 4.4 fb respectively for $m_{H_1} = 300$ GeV, $700$ GeV and $1$ TeV respectively,  whereas we obtain 32.1, 2.4 and 1.4 fb with our extrapolation procedure and without systematic uncertainties. We find that we approximately 
reproduce\footnote{We however stress that we reproduce the CMS projected 95\% C.L. cross section upper limits without systematic uncertainties (41, 3.4 and 2.4 fb respectively for $m_{H_1} = 300$ GeV, $700$ GeV and $1$ TeV) to about $30 - 40\%$ accuracy.} the CMS projected 95\% C.L. cross section upper limits by assuming a 2\% systematic uncertainty for the MMR benchmarks, $m_{H_1} = 700$ GeV, $1$ TeV, and a 0.1\% systematic uncertainty for the LMR benchmark $m_{H_1} = 300$ GeV.
Hence in our HL-LHC extrapolations (as well as for the HE-LHC and FCC-hh extrapolations from section~\ref{sec:FCChh}) we will consider a flat systematic error of 2\% for the MMR category and of 0.1\% for the LMR category (we also show the HL-LHC, HE-LHC and FCC-hh extrapolations in the absence of systematic uncertainties in Appendix~\ref{app:H1H2H2nosys}), which are also the values adopted in section~\ref{sec:enlarging} for the current upper limits. We stress again that these uncertainties are dominated by the data-driven SM background modeling\footnote{See sections 5.1.2 and 5.1.3 of Ref.~\cite{DiMicco:2019ngk} for a detailed discussion.}.


\begin{figure}[ht!]
\centering

$\vcenter{\hbox{\includegraphics[width=0.75\textwidth]{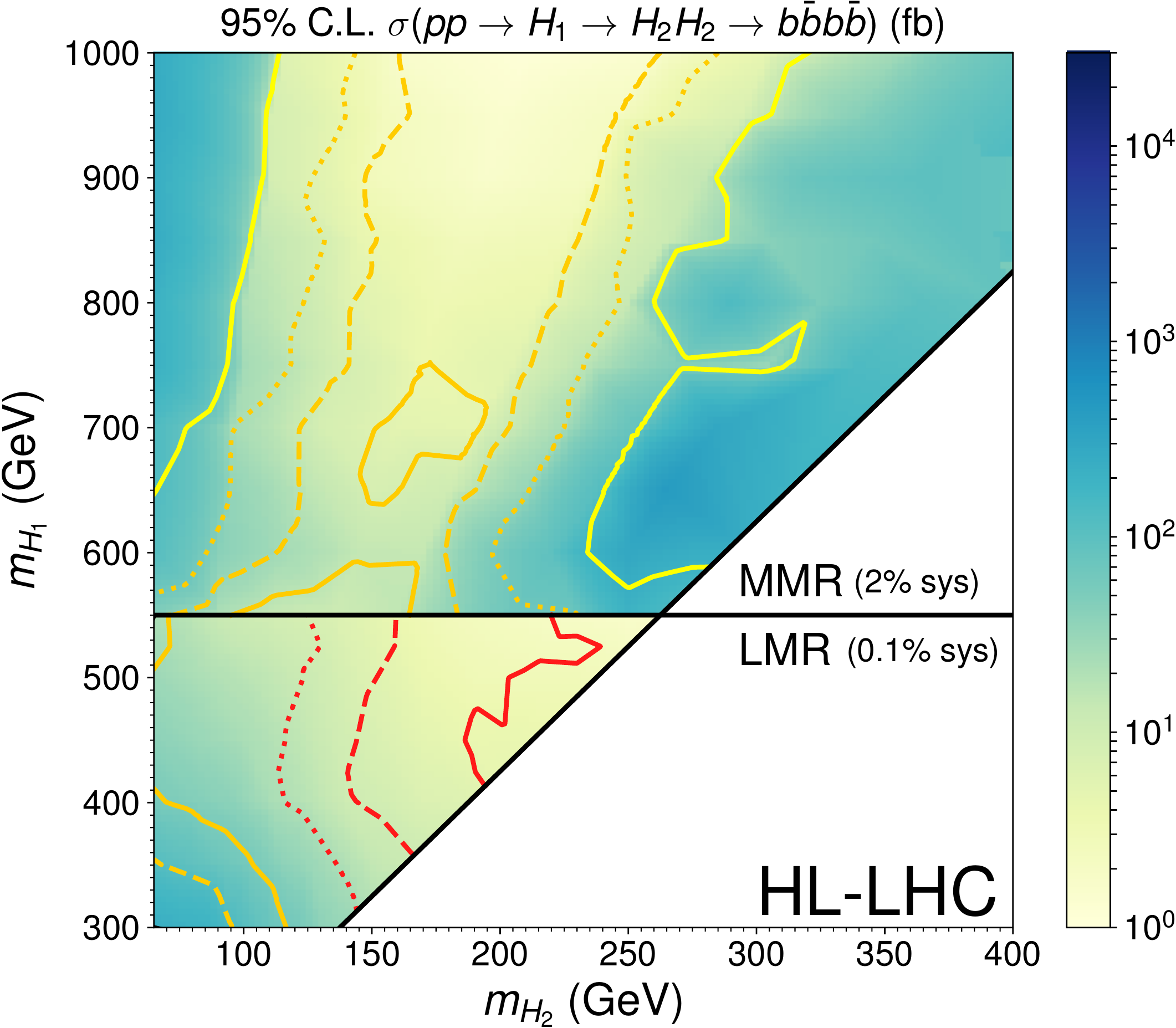}}}$
\hspace{4mm}
$\vcenter{\hbox{\includegraphics[width=0.17\textwidth]{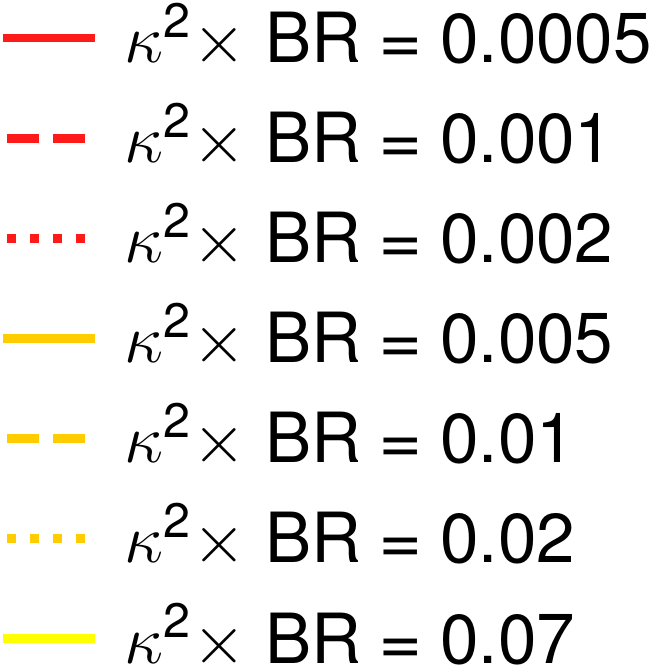}}}$

\caption{\small Projected HL-LHC 95\% C.L. upper limit on the $pp\to H_1 \to H_2 H_2 \to b\bar b b \bar b$ signal cross section (in fb) in the ($m_{H_2}, \, m_{H_1}$) 
plane for the MMR ($m_{H_1} > 550$ GeV, assuming a $2\%$ background systematic~uncertainty) and LMR ($m_{H_1} < 550$ GeV, assuming a $0.1\%$ background systematic~uncertainty) categories. The various contours correspond to the projected 95\% C.L. upper limits on the 
sensitivity $\kappa^2 \times {\rm BR}$ from Eq.~\eqref{eq:kappaBR}, see text for details.}
\label{fig:HL-LHC_limits}
\end{figure}

The projected $95 \%$ C.L upper limits on the signal cross section for HL-LHC are shown in Fig.~\ref{fig:HL-LHC_limits} for both the MMR ($m_{H_1} > 550$ GeV) and LMR categories, together with the 
$\kappa^2 \times \mathrm{BR}$ sensitivity defined in Eq.~\eqref{eq:kappaBR}. By comparing them with the results from section~\ref{sec:enlarging}, we observe a factor $I \simeq 5-10$ improvement in the sensitivity for the HL-LHC with respect to the present reach in the $\kappa^2 \times \mathrm{BR}$ factor for the LMR category, becoming larger as $m_{H_2}$ increases. For the MMR category the improvement is significantly smaller ($I \sim 2-4$) for $m_{H_2} \lesssim 125$ GeV, reaching however $I > 10$ values for $m_{H_2} > 200$ GeV.
Values of $\kappa^2 \times \mathrm{BR}$ and the improvement in sensitivity $I$ for several benchmarks in the ($m_{H_2}$, $m_{H_1}$) plane are given for HL-LHC, HE-LHC and FCC-hh in Table~\ref{Table2}.

\begin{figure}[t!]
\begin{center}
\includegraphics[width=0.48\textwidth]{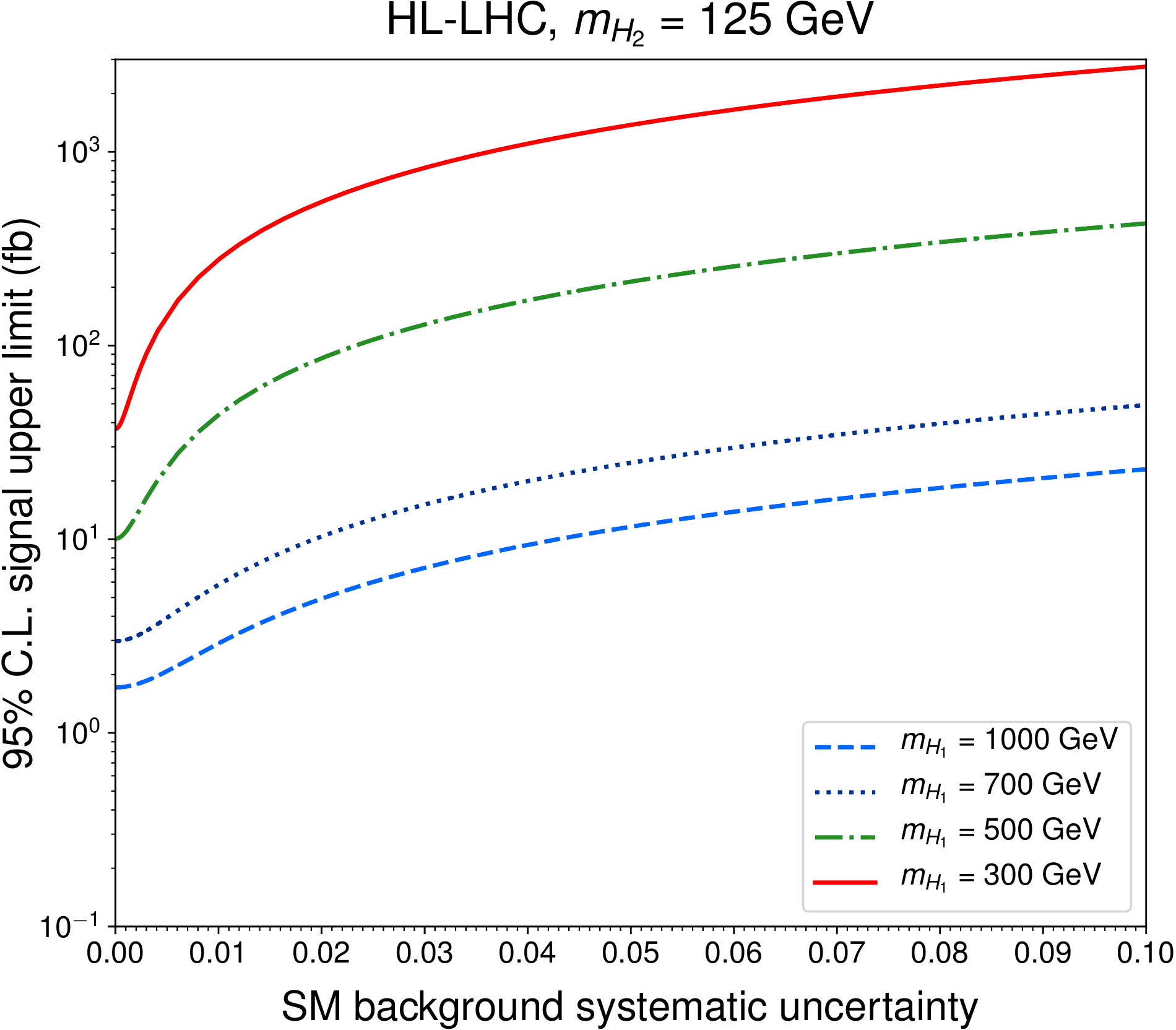}\hfill
\includegraphics[width=0.48\textwidth]{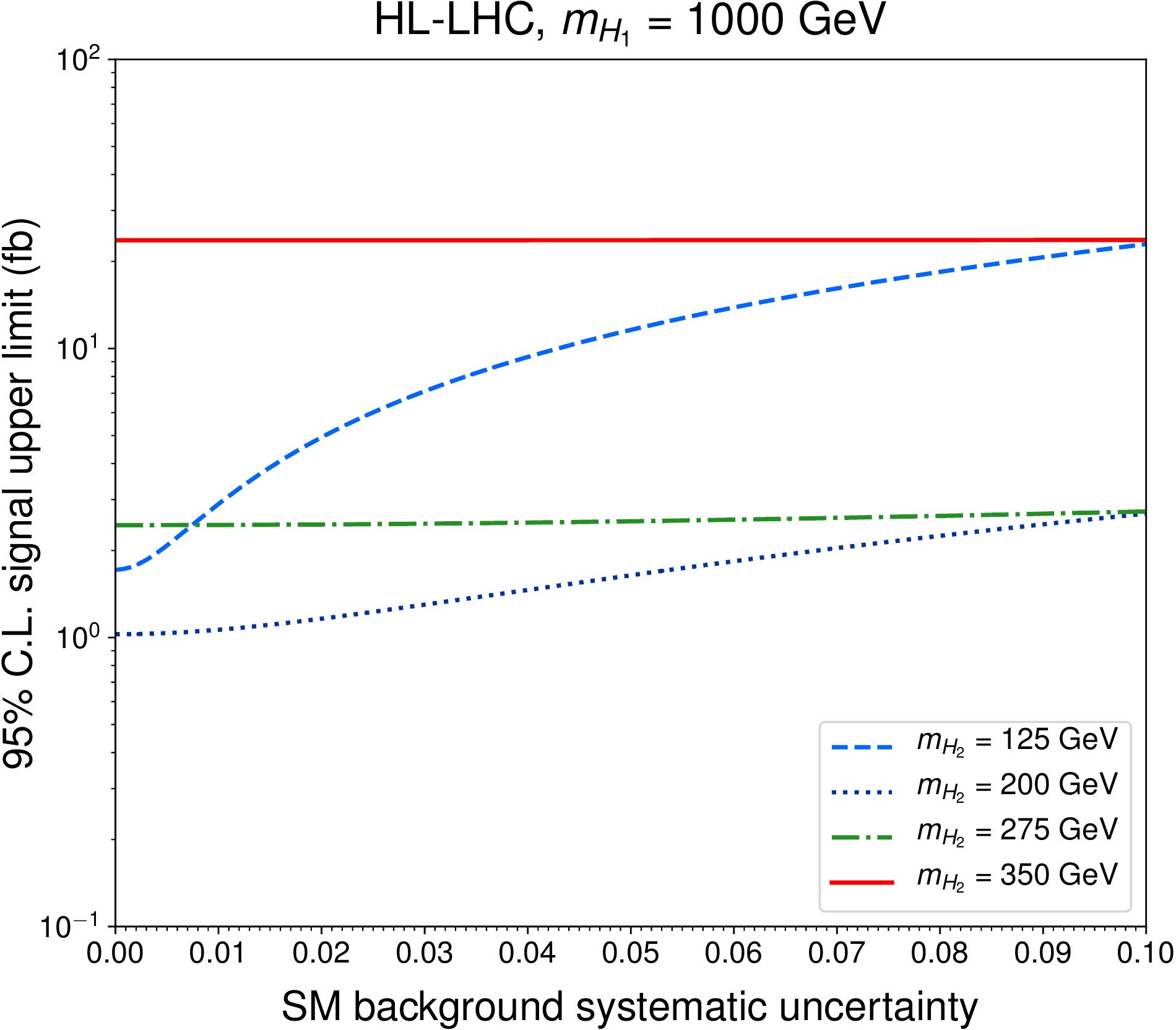}\\
\caption{\small 95\% C.L. upper limit on the $pp\to H_1 \to H_2 H_2 \to b \bar{b} b \bar{b}$ cross section at the HL-LHC as a function of the SM background systematic uncertainty. We fix $\mhone$ $(\mhtwo)$ in the left (right) panel.}
\label{fig:syst_HL-LHC}
\end{center}

\vspace{-4mm}

\end{figure}

\vspace{1mm}

It is worth studying here in more detail the impact of systematic errors on the projected HL-LHC $95\%$ C.L. signal upper limits, since 
in this case (as opposed to that of section~\ref{sec:enlarging}) the upper limits start to become systematics dominated. 
For the discussion of systematic errors we follow Ref.~\cite{HLHELHCCommonSystematics}. Considering that we are dealing mostly with $b$-jets (with a reported 2 -- 6\% overall systematic uncertainty), and taking into account additional ${\cal O}(1 \%)$ sources of systematics ({\emph{e.g}}~integrated luminosity, jet energy scale...) treated as uncorrelated, we may expect a 1 -- 10\% range for the overall systematic uncertainty. We nevertheless stress that due to the large statistics for the data-driven SM background, background systematic uncertainties are expected to be small. In Fig.~\ref{fig:syst_HL-LHC} we show the variation of the projected HL-LHC $95\%$ C.L. upper limit on the signal cross section as a function of the SM background systematic uncertainty, for fixed $\mhtwo = 125~\rm{GeV}$ (left panel) and fixed $\mhone = 1~\rm{TeV}$ (right panel). %
We observe how increasing the SM background systematic error leads to a saturation of the 95\% C.L. upper limit on the signal cross section when systematic uncertainties dominate over statistical ones. In the left panel of Fig.~\ref{fig:syst_HL-LHC} we show that for low $m_{H_2}$, where the multi-jet SM background is large, already a few \% systematic error on the SM background leads to an important increase of the 95\% C.L. signal upper limit, thus stressing the key importance of having systematic effects under control for this study. 
At the same time, increasing the mass $m_{H_2}$ leads to a significant decrease of the SM background yield in the SR and thus to a smaller impact of the corresponding background systematics, as can be seen from 
the right panel of Fig.~\ref{fig:syst_HL-LHC}. This panel also shows that for the MMR category (in this case, for a benchmark $m_{H_1} = 1$ TeV), the interplay between the decrease of SM background yield and the decrease of signal efficiency in the SR as $m_{H_2}$ increases leads to 
a minimum for the 95\% C.L. signal upper limit (as a function of $m_{H_2}$) for $m_{H_2}$ in the range $m_{H_2} \sim 140$ -- $200$ GeV (recall the discussion at the end of section~\ref{sec:enlarging}). 

\begin{table}[t!]
\begin{center}
\begin{tabular}{|C{0.6cm}|C{0.6cm}|C{2.5cm}|C{1.7cm}|C{0.75cm}|C{1.7cm}|C{0.75cm}|C{1.7cm}|C{0.75cm}|}
\cline{3-9}
\multicolumn{2}{c|}{ } & LHC 35.9 fb$^{-1}$  & \multicolumn{2}{c|}{HL-LHC}  & \multicolumn{2}{c|}{HE-LHC} & \multicolumn{2}{c|}{FCC-hh}  \\ \hline
$m_{H_2}$ & $m_{H_1}$ & $\kappa\times$BR & $\kappa\times$BR & $I$ & $\kappa\times$BR & $I$ & $\kappa\times$BR & $I$  \\ \hline
\multirow{4}{*}{75}
&  300  &  $7.1\times 10^{-2}$  &  $1.6\times 10^{-2}$  &  4.5  &  $1.4\times 10^{-2}$  &  5.0  &  $1.4\times 10^{-2}$  &  5.0 \\ \cline{2-9}
&  500  & $2.8\times 10^{-2}$ & $4.2\times 10^{-3}$ &  6.6  & $3.2\times 10^{-3}$ &  8.7  & $3.2\times 10^{-3}$ &  8.8 \\ \cline{2-9}
&  700  & $1.1\times 10^{-1}$ & $7.7\times 10^{-2}$ &  1.5  & $7.7\times 10^{-2}$ &  1.5  & $7.7\times 10^{-2}$ &  1.5 \\ \cline{2-9}
&  900  & $9.5\times 10^{-1}$ & $5.8\times 10^{-1}$ &  1.6  & $5.8\times 10^{-1}$ &  1.6  & $5.8\times 10^{-1}$ &  1.6\\ \hline
\multirow{4}{*}{125}
&  300  & $3.0\times 10^{-2}$ & $3.7\times 10^{-3}$ &  8.0  & $2.3\times 10^{-3}$ &  12.8  & $2.2\times 10^{-3}$ &  13.5 \\ \cline{2-9}
&  500  & $1.8\times 10^{-2}$ & $1.9\times 10^{-3}$ &  9.4  & $8.5\times 10^{-4}$ &  21.3  & $7.6\times 10^{-4}$ &  23.8 \\ \cline{2-9}
&  700  & $2.9\times 10^{-2}$ & $8.7\times 10^{-3}$ &  3.3  & $8.3\times 10^{-3}$ &  3.4  & $8.3\times 10^{-3}$ &  3.4 \\ \cline{2-9}
&  900  & $7.7\times 10^{-2}$ & $2.1\times 10^{-2}$ &  3.7  & $2.0\times 10^{-2}$ &  3.9  & $1.9\times 10^{-2}$ &  4.0 \\ \hline
\multirow{3}{*}{175}
&  500  & $6.7\times 10^{-3}$ & $6.3\times 10^{-4}$ &  10.6  & $1.6\times 10^{-4}$ &  41.9  & $8.6\times 10^{-5}$ &  77.9 \\ \cline{2-9}
&  700  & $3.7\times 10^{-2}$ & $4.1\times 10^{-3}$ &  8.9  & $2.8\times 10^{-3}$ &  13.0  & $2.7\times 10^{-3}$ &  13.3 \\ \cline{2-9}
&  900  & $6.5\times 10^{-2}$ & $7.0\times 10^{-3}$ &  9.6  & $4.5\times 10^{-3}$ &  14.4  & $4.4\times 10^{-3}$ &  14.8 \\ \hline
\multirow{3}{*}{225}
&  500  & $5.9\times 10^{-3}$ & $4.9\times 10^{-4}$ &  12.0  & $1.1\times 10^{-4}$ &  53.6  & $3.4\times 10^{-5}$ &  174 \\ \cline{2-9}
&  700  & $2.8\times 10^{-1}$ & $1.9\times 10^{-2}$ &  14.5  & $7.7\times 10^{-3}$ &  36.3  & $6.8\times 10^{-3}$ &  40.8 \\ \cline{2-9}
&  900  & $1.2\times 10^{-1}$ & $6.7\times 10^{-3}$ &  17.7  & $2.2\times 10^{-3}$ &  54.1  & $1.9\times 10^{-3}$ &  64.0 \\ \hline
\multirow{2}{*}{275}
&  700  &$5.0\times 10^{0\phantom{-}}$  & $1.8\times 10^{-1}$ &  27.2  & $4.2\times 10^{-2}$ &  121  & $2.5\times 10^{-2}$ &  203 \\ \cline{2-9}
&  900  & $1.3\times 10^{0\phantom{-}}$  & $4.4\times 10^{-2}$ &  29.5  & $8.9\times 10^{-3}$ &  145  & $5.1\times 10^{-3}$ &  253 \\ \hline
\multirow{2}{*}{325}
&  700  &  $6.4\times 10^{0\phantom{-}}$  & $1.4\times 10^{-1}$ &  45.1  & $2.5\times 10^{-2}$ &  258  & $9.5\times 10^{-3}$ &  670 \\ \cline{2-9}
&  900  & $1.6\times 10^{1\phantom{-}}$ & $2.6\times 10^{-1}$ &  59.4  & $3.7\times 10^{-2}$ &  422  & $1.1\times 10^{-2}$ &  1445 \\ \hline
\end{tabular}
\caption{\small Value of $\kappa^2 \times$BR for different $m_{H_2}$, $m_{H_1}$ (in GeV) benchmarks, for our proposed search with current ($\sqrt{s} = 13$ TeV, 35.9 fb$^{-1}$) LHC data (assuming respectively a 2\% and 0.1\% systematic uncertainty for the MMR and LMR benchmarks), as well as the extrapolations to HL-LHC, HE-LHC and FCC-hh (assuming the same uncertainties). We also give the sensitivity improvement $I$ for HL-LHC, HE-LHC and FCC-hh.} 
\label{Table2}
\end{center}

\vspace{-3mm}

\end{table}

We close this section with a few remarks on the impact of $b$-tagging and trigger efficiency. We first note that accounting for potential improvements on $b$-tagging for the HL-LHC and future colliders in our projections is rather difficult due to: {\sl i)} the $b$-tagging efficiency being a phase-space dependent ($p_T, \eta$) quantity. This is already an important issue for the current analysis, as detailed in Appendix~\ref{app:btagging}; {\sl ii)} our lack of knowledge of the $b$-jet truth content of the SM multi-jet background; {\sl iii)} noting that improvements would come from a deep-learning algorithm which will most likely not yield a flat rescaling in phase-space; {\sl iv)} noting that improvements strongly depend on both the detector capabilities and their performance, as well as on the pileup conditions, all of which are not fully known for future colliders like HE-LHC and FCC-hh. Nevertheless, we can obtain a rough idea of potential improvements quoted in the recent literature. On one hand, the effect of changes in $b$-tagging efficiency on the overall signal strength uncertainty has been evaluated by the CMS collaboration, showing that an improvement of 10\% in the $b$-tagging efficiency leads to a relative improvement in the signal strength uncertainty of up to 6\%~\cite{CMS:2018qgz}. On the other hand, the inclusion of timing information (which helps to reduce the number of spurious reconstructed secondary vertices by $\sim 30$\%) provides an increase in the $b$-tagging efficiency of about 4 -- 6\% depending on the pseudorapidity, evaluated for the same mis-tag rate~\cite{Collaboration:2296612}. Finally, while the challenging data-taking conditions at the HL-LHC could worsen the $b$-tagging efficiency, the new inner tracker detector as well as novel reconstruction techniques could provide a sizeable improvement. For example, it has been estimated that the upgrades of the inner tracker would lead to an 8\% improvement in efficiency~\cite{Collaboration:2017mtb}.

\begin{figure}[h!]
\centering

$\vcenter{\hbox{\includegraphics[width=0.75\textwidth]{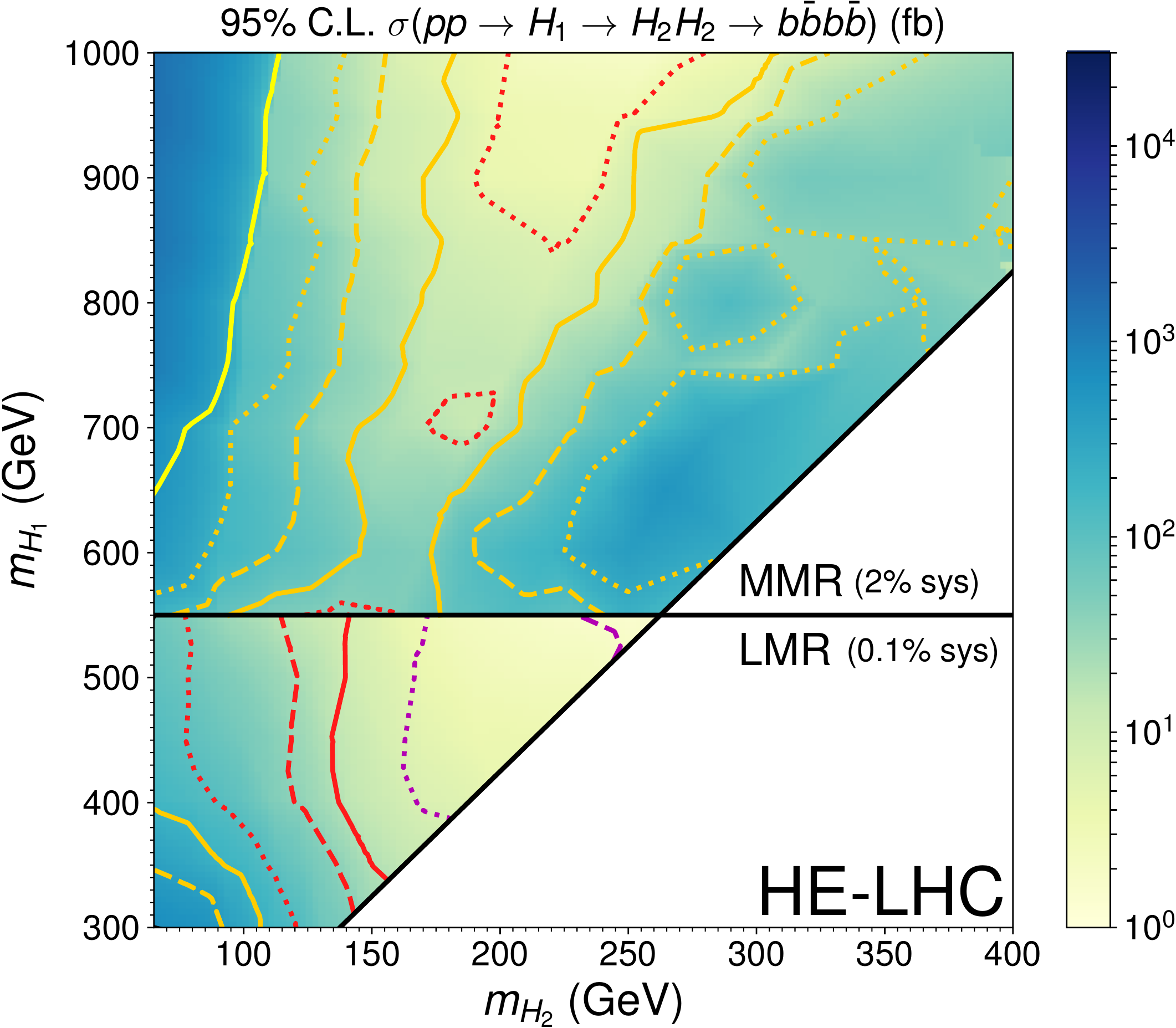}}}$
\hspace{4mm}
$\vcenter{\hbox{\includegraphics[width=0.17\textwidth]{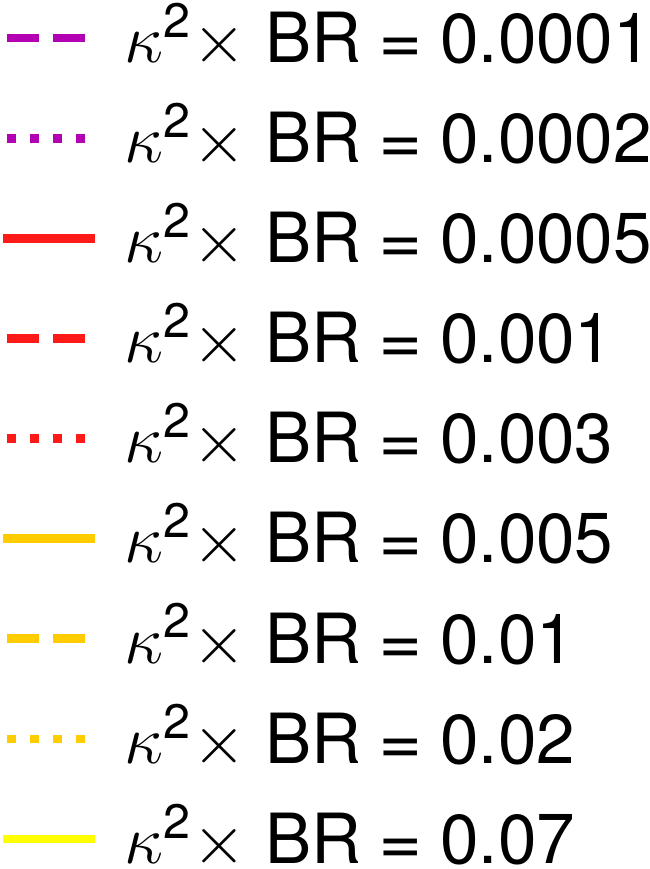}}}$

\caption{\small Projected HE-LHC 95\% C.L. upper limit on the $pp\to H_1 \to H_2 H_2 \to b\bar b b \bar b$ signal cross section (in fb) in the ($m_{H_2}, \, m_{H_1}$) 
plane for the MMR ($m_{H_1} > 550$ GeV, assuming a $2\%$ background systematic~uncertainty) and LMR ($m_{H_1} < 550$ GeV, assuming a $0.1\%$ background systematic~uncertainty) categories. The various contours correspond to the projected 95\% C.L. upper limits on the 
sensitivity $\kappa^2 \times {\rm BR}$ from Eq.~\eqref{eq:kappaBR}, see text for details.}
\label{fig:HE-LHC_limits}

\vspace{-2mm}

\end{figure}

\begin{figure}[t!]
\centering

$\vcenter{\hbox{\includegraphics[width=0.75\textwidth]{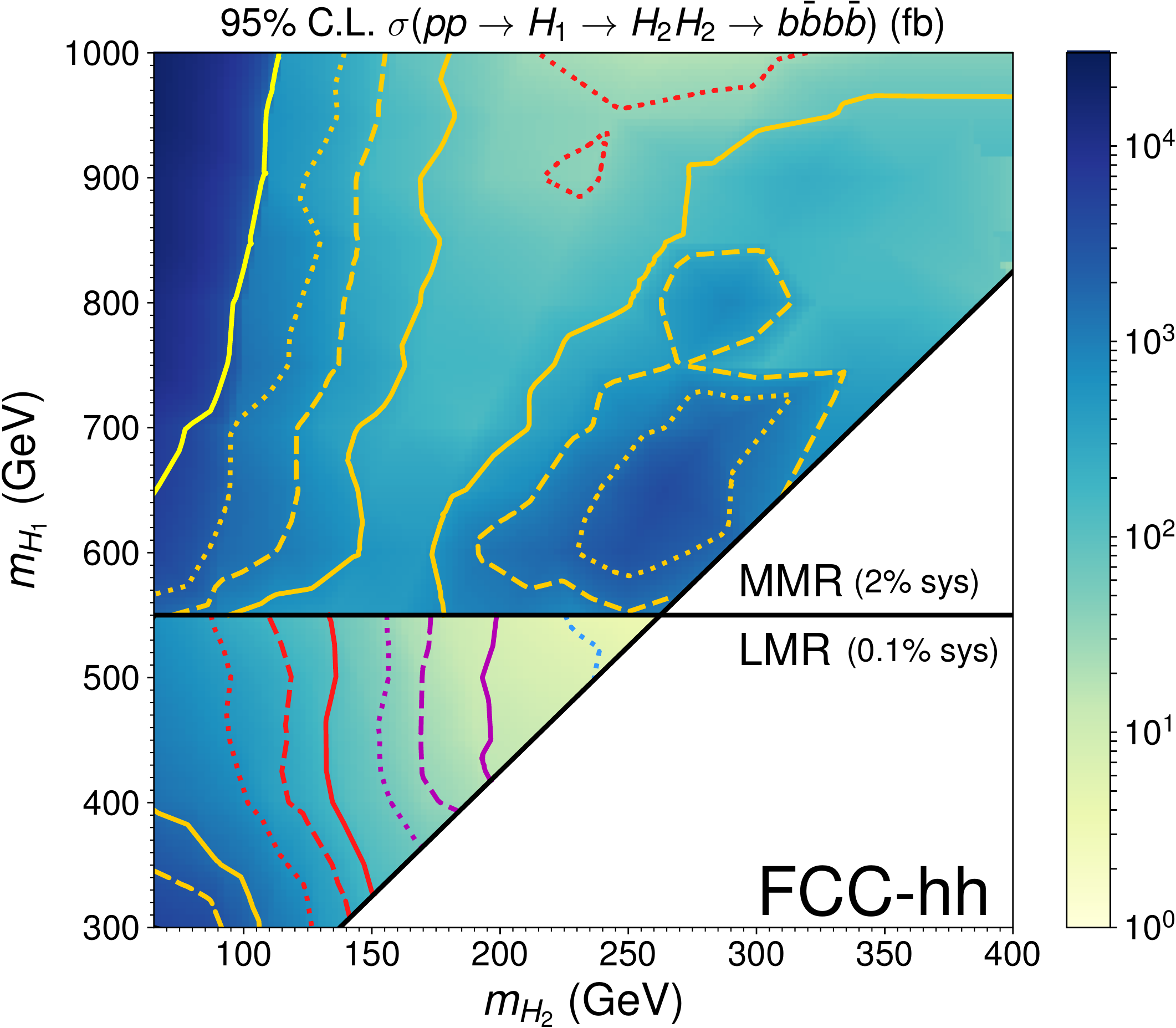}}}$
\hspace{4mm}
$\vcenter{\hbox{\includegraphics[width=0.18\textwidth]{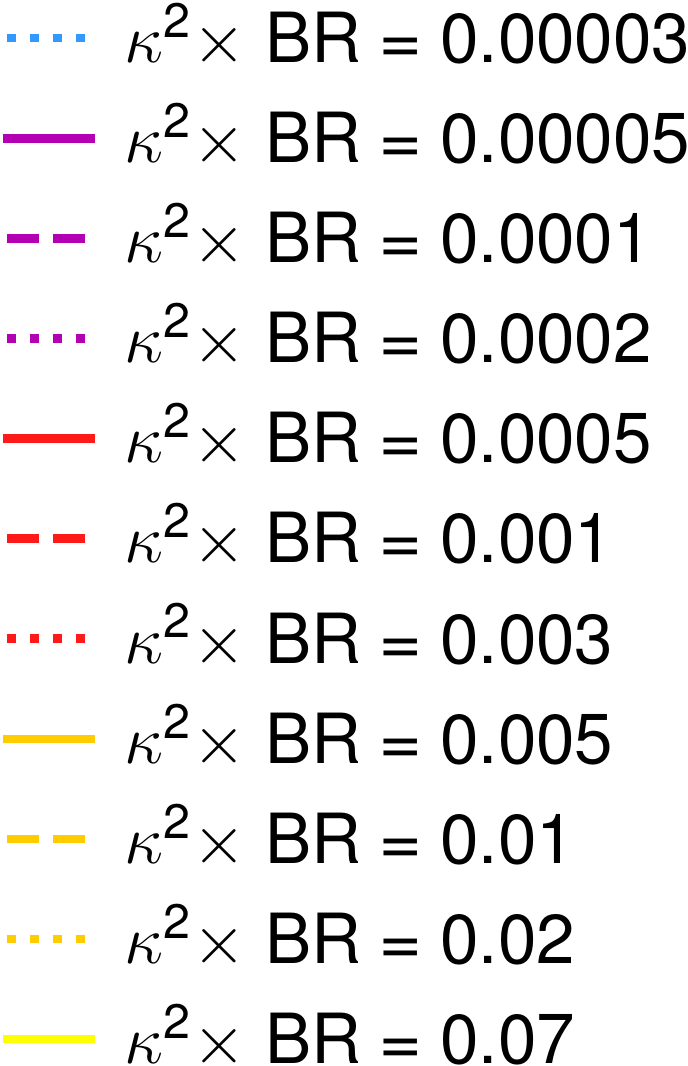}}}$

\caption{\small Projected FCC-hh 95\% C.L. upper limit on the $pp\to H_1 \to H_2 H_2 \to b\bar b b \bar b$ signal cross section (in fb) in the ($m_{H_2}, \, m_{H_1}$) 
plane for the MMR ($m_{H_1} > 550$ GeV, assuming a $2\%$ background systematic~uncertainty) and LMR ($m_{H_1} < 550$ GeV, assuming a $0.1\%$ background systematic~uncertainty) categories. The various contours correspond to the projected 95\% C.L. upper limits on the 
sensitivity $\kappa^2 \times {\rm BR}$ from Eq.~\eqref{eq:kappaBR}, see text for details.}
\label{fig:FCC-hh_limits}

\vspace{-4mm}

\end{figure}

\subsection{Extrapolation to HE-LHC and FCC-hh}
\label{sec:FCChh}

Using the results from section~\ref{sec:HL_LHC_1}, here we provide an extrapolation of the 95\% C.L. $\sigma(p p \to H_1 \to H_2 H_2 \to b \bar{b} b \bar{b})$ upper limits to a $\sqrt{s} = 27$ TeV HE-LHC collider and a $\sqrt{s} = 100$ TeV FCC-hh collider, see Table~\ref{tab:futcoll}. As for the extrapolation to HL-LHC performed in the previous section, we assume here a background systematic error of 0.1\% and 2\% for the LMR and MMR categories, respectively (the results for HE-LHC and FCC-hh without background systematic uncertainties are shown in Appendix~\ref{app:H1H2H2nosys}). Our results for HE-LHC and FCC-hh are shown respectively in Fig.~\ref{fig:HE-LHC_limits} and Fig.~\ref{fig:FCC-hh_limits}. 
By comparing these with the HL-LHC results from Fig.~\ref{fig:HL-LHC_limits}, we can readily see that SM background systematic uncertainties significantly hinder the potential improvement in sensitivity of HE-LHC and FCC-hh for the MMR category. The 2$\%$ SM background systematics result in the signal sensitivity being dominantly driven by this error, such that the much larger integrated luminosities of the HE-LHC and the FCC-hh with respect to the HL-LHC do not significantly increase the sensitivity to our signal, except for large $m_{H_2}$. There the acceptances for the MMR category are rather low, and the limits remain statistics dominated at the HL-LHC and HE-LHC (and even for FCC-hh for the largest values of $m_{H_2}$). For the LMR category (with 0.1$\%$ SM background systematics) the limits are still systematics dominated for $m_{H_2} \lesssim 150$ GeV and the HE-LHC/FCC-hh do not provide a large improvement in signal sensitivity. In contrast, for $m_{H_2} \in [150,\,250]\;$GeV and $m_{H_1} < 550$ GeV there is a major improvement in signal sensitivity for HE-LHC and particularly for FCC-hh. In Table~\ref{Table2} we provide the value of $\kappa^2 \times$ BR as well as the sensitivity improvement $I$ from Eq.~\eqref{eq:sensitivity_resc} for specific 
($m_{H_2}$, $m_{H_1}$) benchmarks. Overall, we observe that HE-LHC and FCC-hh yield a big improvement in sensitivity (given the SM background systematics assumed in this work) only for large values of $m_{H_2}$ (within both LMR and MMR categories), where the SM background yield is suppressed and the search is limited by background statistics rather than systematics. However, we stress that a reduction of SM background systematic uncertainties could result in a large sensitivity improvement for HE-LHC and FCC-hh also for smaller values of $m_{H_2}$, as the results of Appendix~\ref{app:H1H2H2nosys} clearly show.

\section{$H_1 \to H_2 H_2$ as a probe of extended Higgs sectors}
\label{sec:models}

We now analyze the LHC sensitivity of the proposed search $p p \to H_1 \to H_2 H_2 \to b \bar{b} b \bar{b}$ in the context of specific extensions of the SM. Our aim here is two-fold. First, we will assess the reach of this search within the parameter space of several well-studied BSM models. Second, we will compare the projected sensitivity of the search to other analyses for BSM scalars, studying their complementarity and identifying where $p p \to H_1 \to H_2 H_2 \to b \bar{b} b \bar{b}$ provides the leading probe of the existence of these new scalars. 
For our analysis we consider three benchmark models: 

\begin{itemize}
\item A simplified model with two real scalar singlets added to the SM in section~\ref{Sec:2_Singlets}, which allows for a direct mapping of the sensitivities derived in sections~\ref{sec:enlarging} and~\ref{sec:enlarging_2} to many BSM scenarios.

\item A Type-I 2HDM scenario in section~\ref{Sec:2HDM}.

\item A 2HDM (of Type-I) with the addition of a real (pseudo)scalar singlet, which captures the features of more complicated scalar sectors as {\emph{e.g.}}~the one of the NMSSM in section~\ref{Sec:2HDM_Singlet}.

\end{itemize}

\subsection{Two singlet scalar extension of the SM}
\label{Sec:2_Singlets}

One of the simplest possibilities is to consider that both the $H_1$ and $H_2$ states come from singlet scalar fields $S_{1,2}$ (see e.g.~\cite{Robens:2019kga} for a recent phenomenological analysis of this scenario). This is a simplified framework to which more complicated models could be mapped. 

The most general scalar potential for the SM Higgs $H$ and two singlet scalars $S_1$, $S_2$ has the following form
\be
\label{Pot_Singlet_12}
\lambda_{a,b,c}\, S_1^a~S_2^b~\left(\left| H \right|^2 - v^2/2\right)^{c}
\ee
with $2 \leq  a+b+2c \leq 4$ (we disregard tadpole terms for $S_{1,2}$). While the most general potential from Eq.~\eqref{Pot_Singlet_12} has 17 free parameters (once the SM Higgs vev $v$ and the Higgs mass $m_h$ are fixed), in practice 
most of them are phenomenologically unimportant and may be safely ignored in the present analysis. In particular, considering the process\footnote{With some abuse of notation, we label for the rest of this section the singlet-like scalar mass eigenstates as $S_1$ and $S_2$ (even if they do not correspond exactly to the singlet scalar states from~\eqref{Pot_Singlet_12} due to singlet-doublet mixing).} $p p \to S_1 \to S_2 S_2 \to b\bar b b\bar b$ we only care about the $ggS_1$, $S_1 S_2 S_2$ and $S_2 b \bar{b}$ interactions. Regarding the former (effective) coupling between $S_1$ and the gluons, this would naively come from the first singlet mixing with the SM Higgs, the mixing given by $\mathrm{sin}\,\alpha_1$. Yet, there are other possibilities, {\emph{e.g}~the additional presence of vector like quarks at/above the TeV scale which couple to $S_1$. These latter interactions allows for having the $gg S_1$
coupling $C_g^1$ as a free parameter in our setup, which can then be traded for the production cross section $\sigma (p p \to S_1)$. Nevertheless, we also discuss below the interplay of our analysis with other LHC searches for $S_1$ when its production at the LHC comes purely from the singlet-doublet mixing $\mathrm{sin}\,\alpha_1$. 

\begin{figure}[h]

\centering

$\vcenter{\hbox{\includegraphics[width=0.46\textwidth]{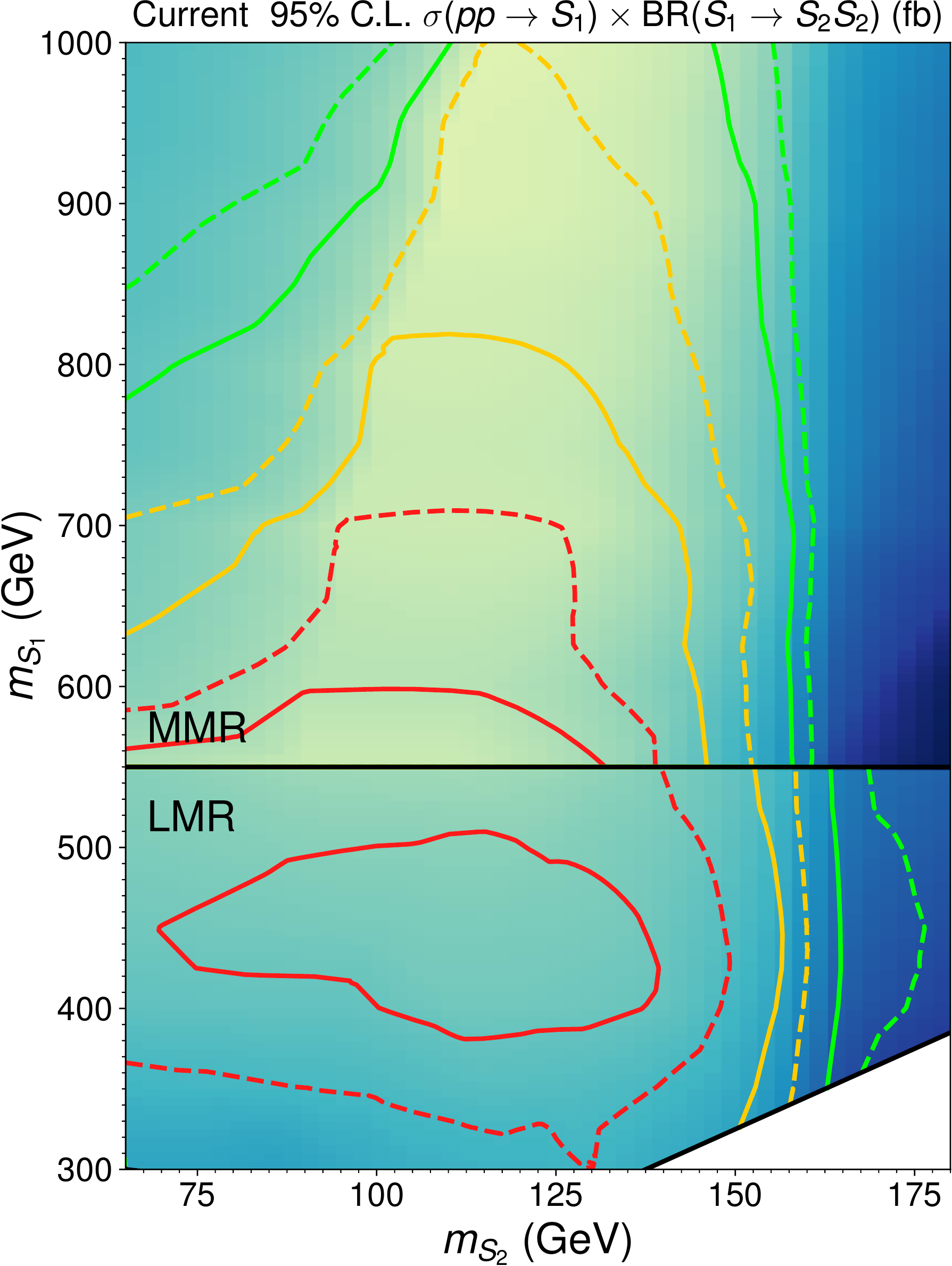}}}$
\hspace{.5mm}
$\vcenter{\hbox{\includegraphics[width=0.52\textwidth]{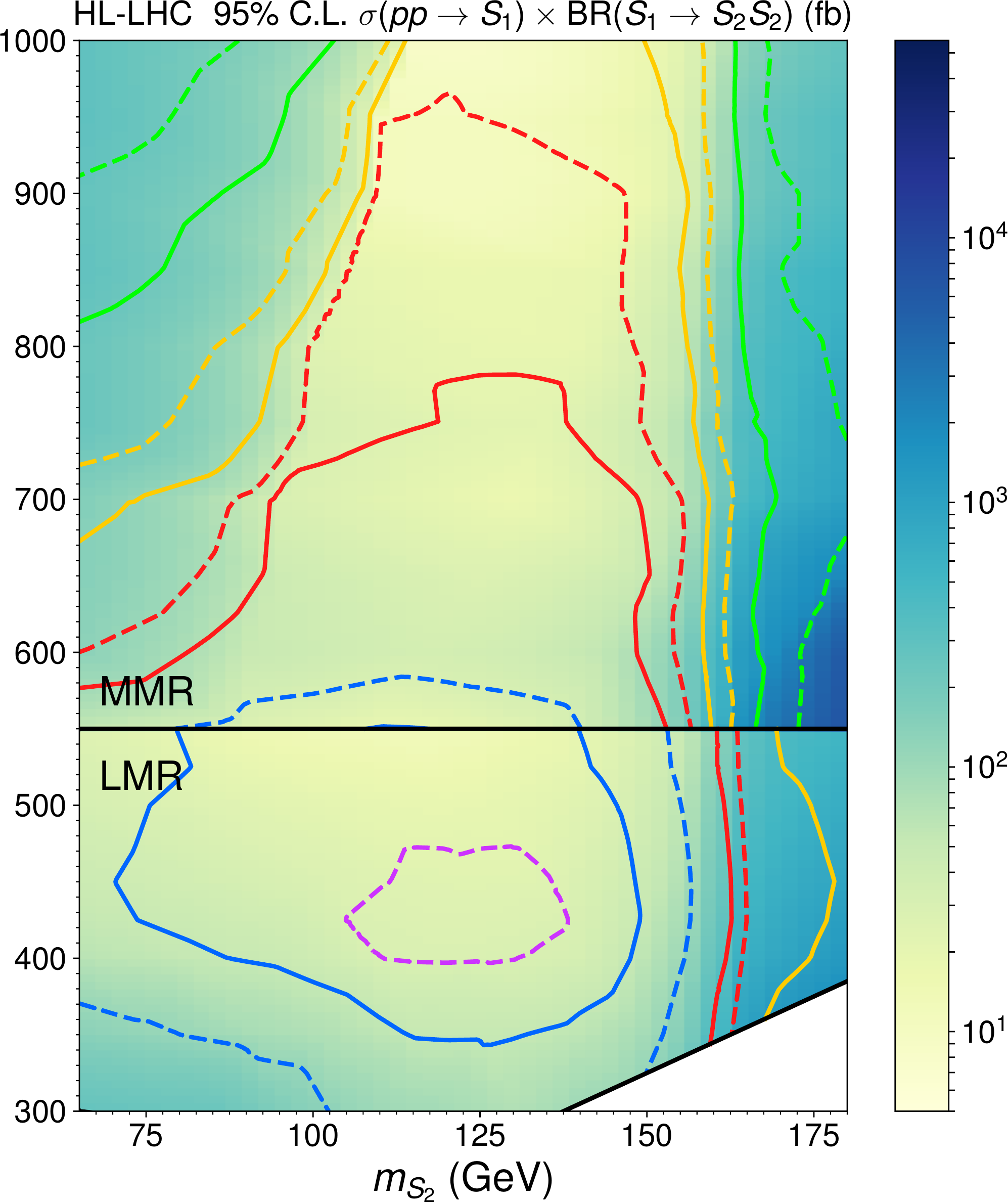}}}$

\vspace{2mm}

$\vcenter{\hbox{\includegraphics[width=0.55\textwidth]{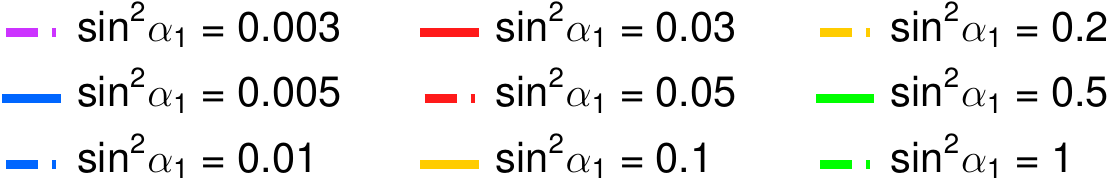}}}$

\caption{\small Present ($\sqrt{s} = 13$ TeV, $35.9$ fb$^{-1}$, left panel) and HL-LHC ($\sqrt{s} = 14$ TeV, $3$ ab$^{-1}$, right panel) 95\% C.L. sensitivity to the $pp\to S_1 \to S_2 S_2$ cross section (in fb) for $m_{S_1} < 550\;$GeV (LMR category) and $m_{S_1} > 550$ GeV (MMR category). Also shown are contours of 95\% C.L. current upper limits (left panel) and HL-LHC projections (right panel) on $\mathrm{sin}^2\alpha_1$ assuming $\mathrm{BR}(S_1 \to S_2 S_2) \simeq 1$.}
\label{fig:LMR_MMR_2Singlets}
\vspace{-2mm}
\end{figure}

The interaction $S_1 S_2 S_2$ is also a free parameter, dominantly controlled by 
the coupling $\lambda_{1,2,0}$ in~\eqref{Pot_Singlet_12} for small singlet-doublet mixing. We note that in this small mixing limit it is possible for the branching fraction BR($S_1 \to S_2 S_2$) to approach unity even with moderate values of $\lambda_{1,2,0}$, strongly suppressing the sensitivity of other search channels for $S_1$ in this case. Finally, the coupling of $S_2$ to the SM fermions is generated via mixing\footnote{In certain scenarios it would also be possible to generate the interaction between $S_2$ and the SM fermions via the dimension $5$ effective operator $(c_q y_q/\Lambda)\, S_2 \bar{Q}_L H q_R$ (see e.g.~\cite{Espinosa:2011eu,Chala:2016ykx}). We however do not consider this possibility here.} 
between the second singlet and the SM Higgs, the mixing given by $\mathrm{sin}\,\alpha_2$. This results in branching fractions of $S_2$ to the SM states equal to those of a SM Higgs boson with mass $m_{S_2}$, independently of the value of  $\mathrm{sin}\,\alpha_2$.
Incidentally, this has the consequence that in this model, our proposed search would be most suited for singlet scalar masses $m_{S_2} \lesssim 150\;$GeV, since for $m_{S_2} \gtrsim 150$ GeV it ceases to be efficient due to the sharp drop in 
BR($S_2 \to b \bar{b}$). A search $p p \to S_1 \to S_2 S_2 \to V V \, V' V'$ (with $V, V^\prime = W,\,Z$) should be most sensitive to the existence of $S_1$ and $S_2$ in this region.

To summarize the above discussion, the present model has as relevant free parameters $C_g^1$, $m_{S_1}$, $m_{S_2}$, BR($S_1 \to S_2 S_2$) and the mixing $\mathrm{sin}\,\alpha_2$. The parameter $C_g^1$ can be traded by either  
the mixing $\mathrm{sin}\,\alpha_1$ or directly the production cross section $\sigma (p p \to S_1)$.  At the same time, the value of $\mathrm{sin}\,\alpha_2$ is only important regarding the complementarity with other collider probes of $S_2$, since BR($S_2 \to b \bar{b}$) is solely determined by $m_{S_2}$.
Using the results from sections~\ref{sec:enlarging} and~\ref{sec:enlarging_2} we show in Fig.~\ref{fig:LMR_MMR_2Singlets}  
the current (left plot) and HL-LHC (right plot) 95\% C.L. exclusion sensitivity to $\sigma (p p \to S_1) \times \mathrm{BR}(S_1 \to S_2 S_2)$ in the mass plane ($m_{S_2},\,m_{S_1}$). In both cases we have assumed a 2\% (0.1\%) SM background systematic uncertainty for $m_{S_1} > 550$ GeV ($m_{S_1} < 550$ GeV), as in sections~\ref{sec:enlarging} and~\ref{sec:HL_LHC_2}. Then, assuming the gluon fusion production of $S_1$ to come exclusively from the Higgs-singlet mixing and setting also $\mathrm{BR}(S_1 \to S_2 S_2) \simeq 1$, the current/projected 95\% C.L. upper limit on $\sigma(p p \to S_1 \to S_2 S_2)$ from Fig.~\ref{fig:LMR_MMR_2Singlets} can be rephrased as an upper limit on $\mathrm{sin}^2\alpha_1$\footnote{We note that $\mathrm{sin}\,\alpha_1 = \kappa$ as defined in Eq.~\eqref{eq:kappaBR}.}, shown as coloured contours in Fig.~\ref{fig:LMR_MMR_2Singlets}. 

\begin{figure}[h]
\begin{center}
\includegraphics[width=0.85\textwidth]{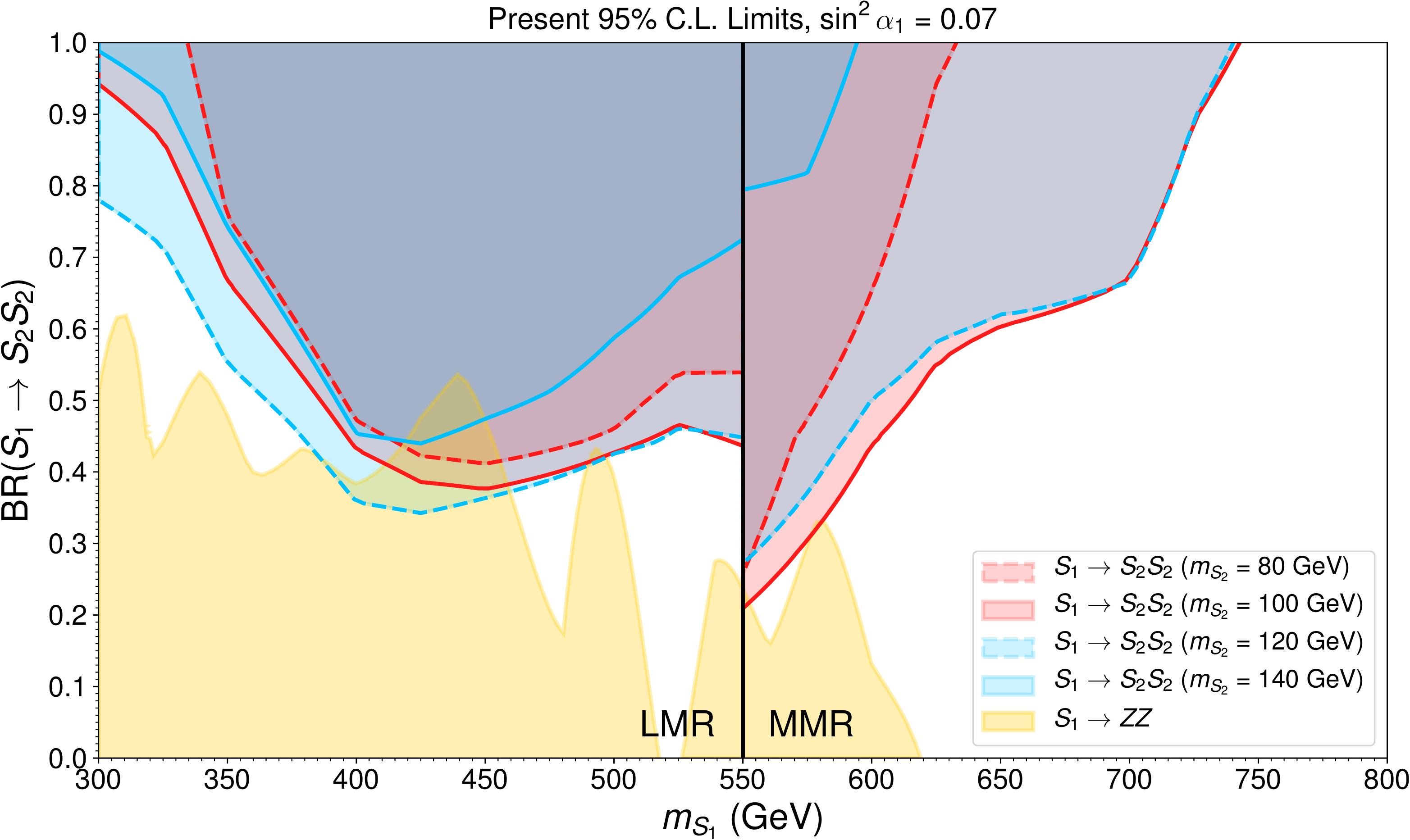}
\caption{\small 95\% C.L. limits on the branching fraction 
$\mathrm{BR}(S_1\to S_2 S_2)$ as a function of $m_{S_1}$ from 
$S_1 \to Z Z$ searches by ATLAS with 36.1 fb$^{-1}$~\cite{Aaboud:2017rel} (yellow region) and from our search $p p \to S_1 \to S_2 S_2 \to b \bar{b} b \bar{b}$ assuming $m_{S_2} = 80$ GeV (solid red), $m_{S_2} = 100$ GeV (dashed red), $m_{S_2} = 120$ GeV (solid blue) and $m_{S_2} = 140$ GeV (dashed blue). The Higgs-singlet mixing has been set to $\mathrm{sin}^2\,\alpha_1 = 0.07$ (satisfying LHC measurements of Higgs signal strengths), and we assume $|\mathrm{sin}\,\alpha_2| \ll |\mathrm{sin}\,\alpha_1|$, $\mathrm{BR}(S_1\to h_{\mathrm{SM}} h_{\mathrm{SM}}) = 0$.}
\label{fig:mixing_2Singlets}
\end{center}


\end{figure}

When the production of $S_1$ at the LHC is due exclusively to the Higgs-singlet mixing $\mathrm{sin}\,\alpha_1$, it is possible to explore the interplay between the $p p \to S_1 \to S_2 S_2 \to b \bar{b} b \bar{b}$ search analyzed in this work, other direct searches for $S_1$ (and $S_2$) and LHC measurements of the Higgs signal strengths. The latter yield the present limit $\mathrm{sin}^2\,\alpha_1 < 0.073$ at 95\% C.L.~(see~\cite{Robens:2019kga}) under the assumption $|\mathrm{sin}\,\alpha_2| \ll |\mathrm{sin}\,\alpha_1|$. In Fig.~\ref{fig:mixing_2Singlets} 
we fix $\mathrm{sin}^2\,\alpha_1 = 0.07$ and show the present sensitivity of our proposed analysis together with the sensitivity of direct BSM Higgs searches in $ZZ$ final states\footnote{Other BSM Higgs searches, e.g. those in $WW$, $\tau\tau$ or $\gamma\gamma$ final states, are significantly less sensitive for the model considered here.} from the latest ATLAS analysis~\cite{Aaboud:2017rel} in the plane ($m_{S_1},\,\mathrm{BR}(S_1\to S_2 S_2)$) for $m_{S_2} = 80,\,100, \,120, \,140$ GeV\footnote{We assume for simplicity 
$\mathrm{BR}(S_1\to h_{\mathrm{SM}} h_{\mathrm{SM}}) = 0$. If this branching fraction is sizable, it will weaken the constraints from Fig.~\ref{fig:mixing_2Singlets}.}. Interestingly, we see that our analysis nicely complements existing searches for BSM scalars, providing a new avenue to probe the singlet-like scalar $S_1$.

\subsection{2HDM}
\label{Sec:2HDM}

The 2HDM represents the simplest scenario where the two BSM states $H_1$ and $H_2$ are contained in one field, a second Higgs doublet. As opposed to the model discussed in section~\ref{Sec:2_Singlets}, for the 2HDM the interactions of $H_{1,2}$ with the SM fermions and gauge bosons are not governed by their mixing with the SM Higgs. 
The scalar potential for a theory with two Higgs doublets $\Phi_{1,2}$ 
(with a softly-broken $\mathbb{Z}_2$-symmetry and no CP-violation) is given by 
\begin{eqnarray}	
\label{2HDM_potential2}
V(\Phi_1,\Phi_2) & = & \mu^2_1 \left|\Phi_1\right|^2 + \mu^2_2\left|\Phi_2\right|^2 - \mu^2 \left[\Phi_1^{\dagger}\Phi_2+\mathrm{h.c.}\right] +\frac{\lambda_1}{2}\left|\Phi_1\right|^4 
+\frac{\lambda_2}{2}\left|\Phi_2\right|^4 \nonumber \\
&  & + \lambda_3 \left|\Phi_1\right|^2\left|\Phi_2\right|^2 +\lambda_4 \left|\Phi_1^{\dagger}\Phi_2\right|^2 +
 \frac{\lambda_5}{2}\left[\left(\Phi_1^{\dagger}\Phi_2\right)^2+\mathrm{h.c.}\right]
\end{eqnarray}
with all scalar potential parameters being real. 
The breaking of EW symmetry is shared between the two doublets, whose vacuum expectation values (vevs) are given by 
$v_{1,2}$ (with $\sqrt{v_1^2 + v_2^2} = v = 246$ GeV, $v_2/v_1 \equiv \mathrm{tan}\, \beta$).
In addition to the 125 GeV Higgs state $h$, the 2HDM scalar sector contains another neutral CP-even scalar $H$, a neutral CP-odd scalar $A$ and a 
charged scalar $H^{\pm}$.
In the following we identify $H$ and $A$ with our neutral BSM states $H_1$ and $H_2$, respectively.

The couplings of the 2HDM scalar states to SM gauge bosons and fermions 
are controlled by $\mathrm{tan}\,\beta$ and by a mixing angle $\alpha$ in the CP-even neutral sector. 
The limit of a SM-like 125 GeV Higgs $h = h_{\mathrm{SM}}$ 
(the so-called ``alignment" limit of the 2HDM~\cite{Gunion:2002zf}) corresponds to $\mathrm{cos}\, (\beta-\alpha) = 0$. 
In addition, there are various possible choices 
for the couplings of the two doublets $\Phi_{1,2}$ to fermions
(see~\cite{Branco:2011iw} for a detailed discussion of fermion couplings in 2HDMs), and in this work we choose for definiteness a so-called Type-I 2HDM.

Given a set of values for $m_{H}$, $m_{A}$, $m_{H^{\pm}}$, $\mathrm{cos}\, (\beta-\alpha)$ and $\mathrm{tan}\,\beta$, 
theoretical constraints dictate 
the allowed range for $\mu^2$ in~\eqref{2HDM_potential2}. 
These constraints are the boundedness from 
below of the 2HDM scalar potential as well as the stability of the EW minimum, and the requirements of unitarity~\cite{Ginzburg:2005dt} and perturbativity on the quartic couplings $\lambda_i$ (see e.g.~\cite{Kling:2016opi,Dorsch:2016tab} for more details).
It is possible that certain choices for $m_{H}$, $m_{A}$, $m_{H^{\pm}}$, $\mathrm{cos}\, (\beta-\alpha)$, $\mathrm{tan}\,\beta$ yield no allowed range for $\mu^2$, making these choices not physically viable.
When a viable range for $\mu^2$ exists, its specific value has an impact on the 2HDM scalar self-couplings (as shown below). 


In the following we consider\footnote{Measurements of EW precision observables require $H^{\pm}$ to be close in mass to either $A$ or $H$, to avoid a large breaking of custodial symmetry. At the same time, bounds from flavour physics constrain $m_{H^{\pm}}$ to be above a certain value at 95\% C.L. (which depends on the 2HDM Type). These motivate our choice $m_{H^{\pm}} \sim m_{H}$.} $m_{H^{\pm}} \simeq m_{H} > 2\, m_{A}$, 
such that the decay $H \to A A$ is open 
and the analysis from sections~\ref{sec:validation} -- \ref{sec:enlarging_2} may be applied to the 2HDM for $H_1 \equiv H$ and $H_2 \equiv A$.
The coupling $g_{H A A}$ is given by~\cite{Kling:2016opi,Dorsch:2016tab} 
\begin{eqnarray}
\label{lambdaHAA}
g_{H A A} = \frac{2}{v} \left[2 \,\left(s_{\beta-\alpha}\frac{c_{2\beta}}{s_{2\beta}} - 
c_{\beta-\alpha} \right) (M^2 - m_{H}^2) - c_{\beta-\alpha}\,(m^2_{H} - 2\, m^2_{A}) \right] \, ,
\end{eqnarray}
with $M^2 \equiv \mu^2/(s_{\beta}\,c_{\beta})$ and we use the notation $c_{\phi} \equiv \mathrm{cos}\,\phi$, $s_{\phi} \equiv \mathrm{sin}\, \phi$. We note that in the alignment limit $c_{\beta-\alpha} = 0$, 
$g_{H A A}$ vanishes for $\mathrm{tan}\,\beta = 1$ and/or $m_{H}^2 = M^2$.
For $m_{H} \gg m_{A}$ the decay $H \to Z A$ will be present together with $H \to A A$, such that both decay modes may compete to be the dominant one 
(for $c_{\beta-\alpha} \neq 0$ the decay modes $H \to W^+ W^-$, $H \to Z Z$ and $H \to h h$ could also be important).
The process $p p \to H \to Z A$ ($Z \to \ell \ell$, $A \to b \bar{b}$) has been searched for by CMS at $\sqrt{s} = 8$ TeV~\cite{Khachatryan:2016are} and by both ATLAS~\cite{Aaboud:2018eoy}\footnote{The 13 TeV search by ATLAS~\cite{Aaboud:2018eoy} interprets its results in terms of an $A \to Z H$ decay, but these are equally applicable to  
$H \to Z A$.} and CMS~\cite{CMS:2016qxc,CMS:2019wml} at $\sqrt{s} = 13$ TeV.
The LHC 13 TeV searches place stringent constraints on the 2HDM parameter space with $m_{H} \gg m_{A}$.

\begin{figure}[h]
\begin{center}
\includegraphics[width=0.495\textwidth]{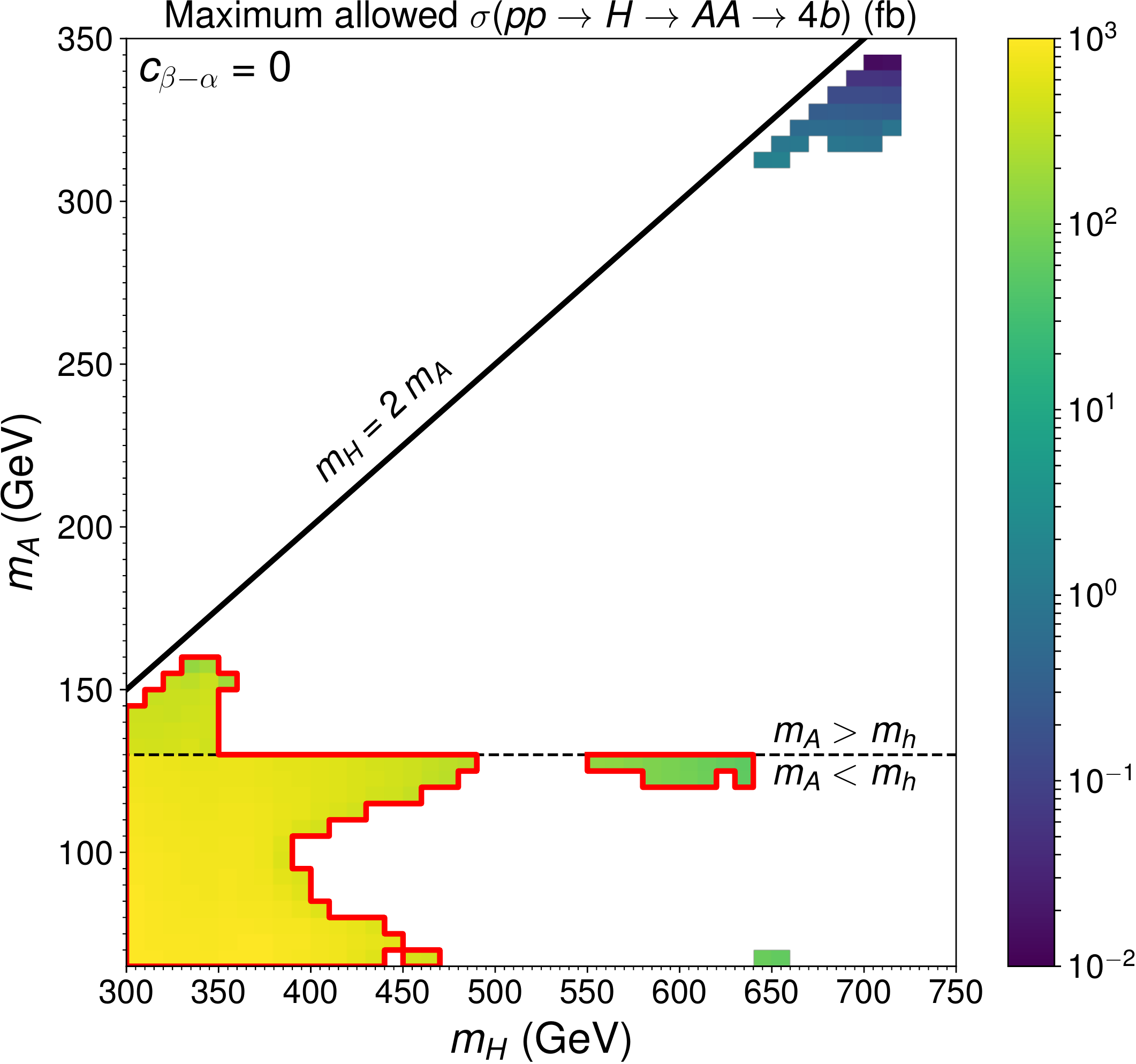}
\includegraphics[width=0.495\textwidth]{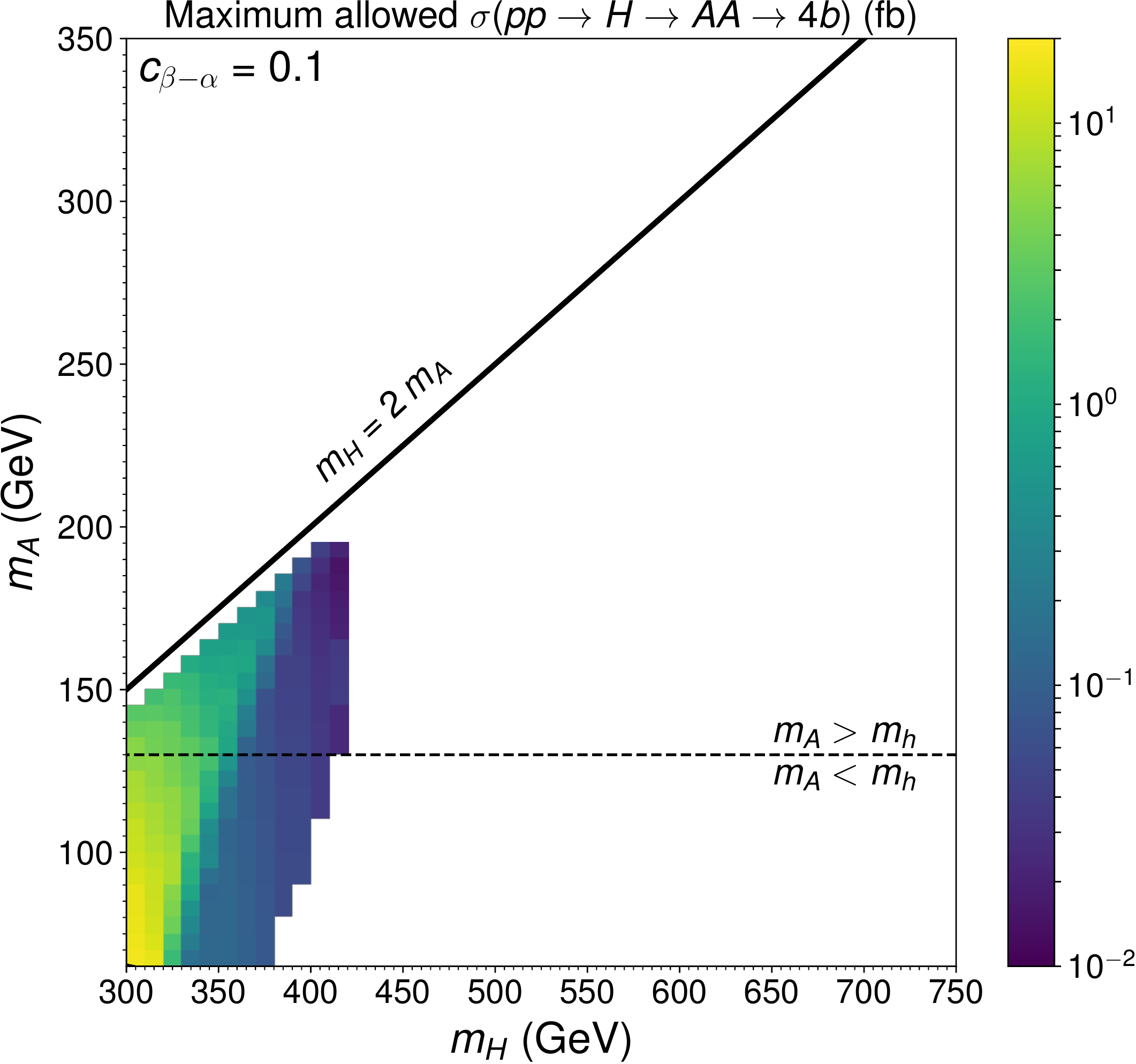}

\caption{\small Maximum allowed LHC 13 TeV cross section $p p \to H \to A A \to b \bar{b} b \bar{b}$ (in fb) as a function of ($m_H$, $m_A$) for $c_{\beta-\alpha} = 0$ (left panel) and $c_{\beta-\alpha} = 0.1$ (right panel), from the combination of 2HDM theoretical constraints and ATLAS/CMS 8 TeV and 13 TeV searches for $p p \to H \to Z A$ ($Z \to \ell \ell$, $A \to b \bar{b}$), $p p \to A/H \to \tau\tau$ and $p p \to H \to Z Z$ (the latter only relevant for $c_{\beta-\alpha} = 0.1$). The region accessible by our proposed search $p p \to H \to A A \to b \bar{b} b \bar{b}$ (recall Figs~\ref{fig:MMR_extension} and~\ref{fig:LMR_extension}) is shown as a red contour.}
\label{fig:2HDM_H_AA_XS}
\end{center}
\vspace{-2mm}
\end{figure}

In order to assess the complementarity between $p p \to H \to Z A$ and $p p \to H \to AA$ searches, 
we perform a scan of the 2HDM parameter space within $m_H \in [300\,\mathrm{GeV}, \,800\,\mathrm{GeV}]$, $m_{H^{\pm}} \in [m_H - 30\,\mathrm{GeV}, \,m_H + 30\,\mathrm{GeV}]$,
$m_A \in [65\,\mathrm{GeV}, \,m_H/2]$ and tan$\,\beta \in [0.5,\,12]$, considering both $c_{\beta-\alpha} = 0$ and $c_{\beta-\alpha} = 0.1$. We use the code~\texttt{2HDMC}~\cite{Eriksson:2009ws} to compute the various branching fractions of the relevant 2HDM states and \texttt{SusHi}~\cite{Harlander:2012pb,Harlander:2016hcx} to obtain their LHC production cross section.
After imposing 2HDM theoretical constraints and ensuring compatibility with the ATLAS/CMS $H \to Z A$ experimental searches, we obtain the maximum possible cross section\footnote{The procedure we follow is, for a given set of values $m_H$, $m_{H^{\pm}}$, $m_A$, $c_{\beta-\alpha}$ and tan$\,\beta$ within our scan, to maximize the BR$(H\to AA)$ within the range of $\mu^2$ allowed by theoretical constraints (which amounts to choosing the value of $M^2$ for which $g_{HAA}$ in~\eqref{lambdaHAA} is maximal in that range), and subsequently discarding the points which are ruled out by the $H \to Z A$ ATLAS/CMS searches.} for the process $p p \to H \to A A \to b \bar{b} b \bar{b}$ in the mass plane ($m_{H}\,, m_A$), shown in Fig.~\ref{fig:2HDM_H_AA_XS} for $c_{\beta-\alpha} = 0$ (left panel) and  $c_{\beta-\alpha} = 0.1$ (right panel). Also included are the latest constraints from $\sqrt{s} = 13$ TeV LHC $p p \to H/A \to \tau\tau$ CMS searches~\cite{Sirunyan:2018zut} and ATLAS di-boson ($p p \to H \to Z Z$) searches~\cite{Aaboud:2017rel} (the latter only relevant for $c_{\beta-\alpha} = 0.1$). In the white region of Fig.~\ref{fig:2HDM_H_AA_XS}, it is not possible to simultaneously satisfy the various theoretical and experimental constraints, while for the coloured region there exist points in our scan that satisfy all constraints, from which we extract a maximum allowed value for $\sigma(p p \to H\to A A \to b \bar{b} b \bar{b})$ as a function of $m_A$ and $m_H$.

As Fig.~\ref{fig:2HDM_H_AA_XS} highlights, the combination of 2HDM theoretical constraints and current experimental limits from BSM Higgs searches rule out a sizable fraction of the ($m_{H}\,, m_A$) parameter space for $m_{H} > 2\, m_{A}$. For $c_{\beta-\alpha} = 0.1$ the search $H \to Z Z$ yields stringent constraints on the 2HDM parameter space, which translate into allowed values $\sigma(p p \to H\to A A \to b \bar{b} b \bar{b}) \lesssim 10$ fb, too low to be probed by our proposed search. In contrast, in the alignment limit $c_{\beta-\alpha} = 0$, the regions that survive the combination of 2HDM theoretical constraints and limits from ATLAS/CMS 13 TeV $H \to Z A$ searches yield a large cross section 
for the process $p p \to H \to A A \to b \bar{b} b \bar{b}$, particularly for $m_A < 130$ GeV.
Fig.~\ref{fig:2HDM_H_AA_XS} shows (red contours) the ($m_{H}\,, m_A$) region for which the current sensitivity of our proposed search allows to probe currently unconstrained 2HDM parameter space, making this search highly complementary to other BSM scalar searches in the 2HDM.

\subsection{2HDM + singlet scalar/pseudoscalar}
\label{Sec:2HDM_Singlet}

We now consider the addition of a real scalar/pseudoscalar singlet field $S$ to the above 2HDM (for a detailed discussion of the 2HDM + complex singlet, see~\cite{Baum:2018zhf}), assuming for simplicity CP conservation in the scalar potential.
This scenario captures the features of richer, more complicated scalar sectors than those analyzed in sections~\ref{Sec:2_Singlets} and~\ref{Sec:2HDM}. At the same time, it is well-motivated as a simplified version of the NMSSM and as a portal to a dark matter sector~\cite{Goncalves:2016iyg,Bauer:2017ota,Abe:2018bpo}. As emphasized in~\cite{Baum:2018zhf}, in this class of extended Higgs sectors the Higgs-to-Higgs cascade decays we discuss in this work are ubiquitous.   

\vspace{1mm}

Besides Eq.~\eqref{2HDM_potential2}, 
the scalar potential for the 2HDM + real singlet field $S$ contains the following terms
\begin{eqnarray}	
\label{2HDM_singlet_potential}
V(S,\Phi_1,\Phi_2) & = & \frac{m_s^2}{2} S^2 + \frac{\lambda_s}{4} S^4 
+ \lambda_{s_1}\, S^2 \left|\Phi_1\right|^2   + \lambda_{s_2}\, S^2 \left|\Phi_2\right|^2
 \nonumber \\
& - & \left[\mu_S\, S\, \Phi_1^{\dagger}\Phi_2+\mathrm{h.c.}\right] +  \mu_{S_1}\, S \left|\Phi_1\right|^2 + \mu_{S_2}\, S \left|\Phi_2\right|^2 + \frac{\mu_3}{3} S^3  \, .
\end{eqnarray}
%
%
We note that if CP is conserved, the last three terms in Eq.~\eqref{2HDM_singlet_potential} are absent for a real pseudoscalar field $S$, and $\mu_S$ has to be purely imaginary.
In the following, we discuss separately the phenomenology for scalar and pseudoscalar $S$, highlighting the differences between them and analyzing the sensitivity of our proposed search in each case.

\vspace{2mm}

\noindent \underline{{\bf Pseudoscalar} $S$}: 

\vspace{1mm}

For $\mu_S \neq 0$ in Eq.~\eqref{2HDM_singlet_potential}, the pseudoscalar $S$ will mix with the 2HDM CP-odd state after EW symmetry breaking, yielding two CP-odd mass eigenstates, which we label $A$, $a$ (with $m_A > m_a$). The lighter mass eigenstate  $a$ is considered here to be mostly singlet-like. In addition to these, the model contains two CP-even scalars $H$ and $h$ (this last one identified with the $125$ GeV Higgs boson) and a charged scalar $H^{\pm}$, as in the 2HDM scenario studied in the previous section. 
Then, assuming for simplicity the 2HDM alignment limit $c_{\beta-\alpha} = 0$, we can analyze the sensitivity of the various LHC probes of the states $H$ and $a$, as well as the interplay among them.

\begin{figure}[h]

\centering

$\vcenter{\hbox{\includegraphics[width=0.495\textwidth]{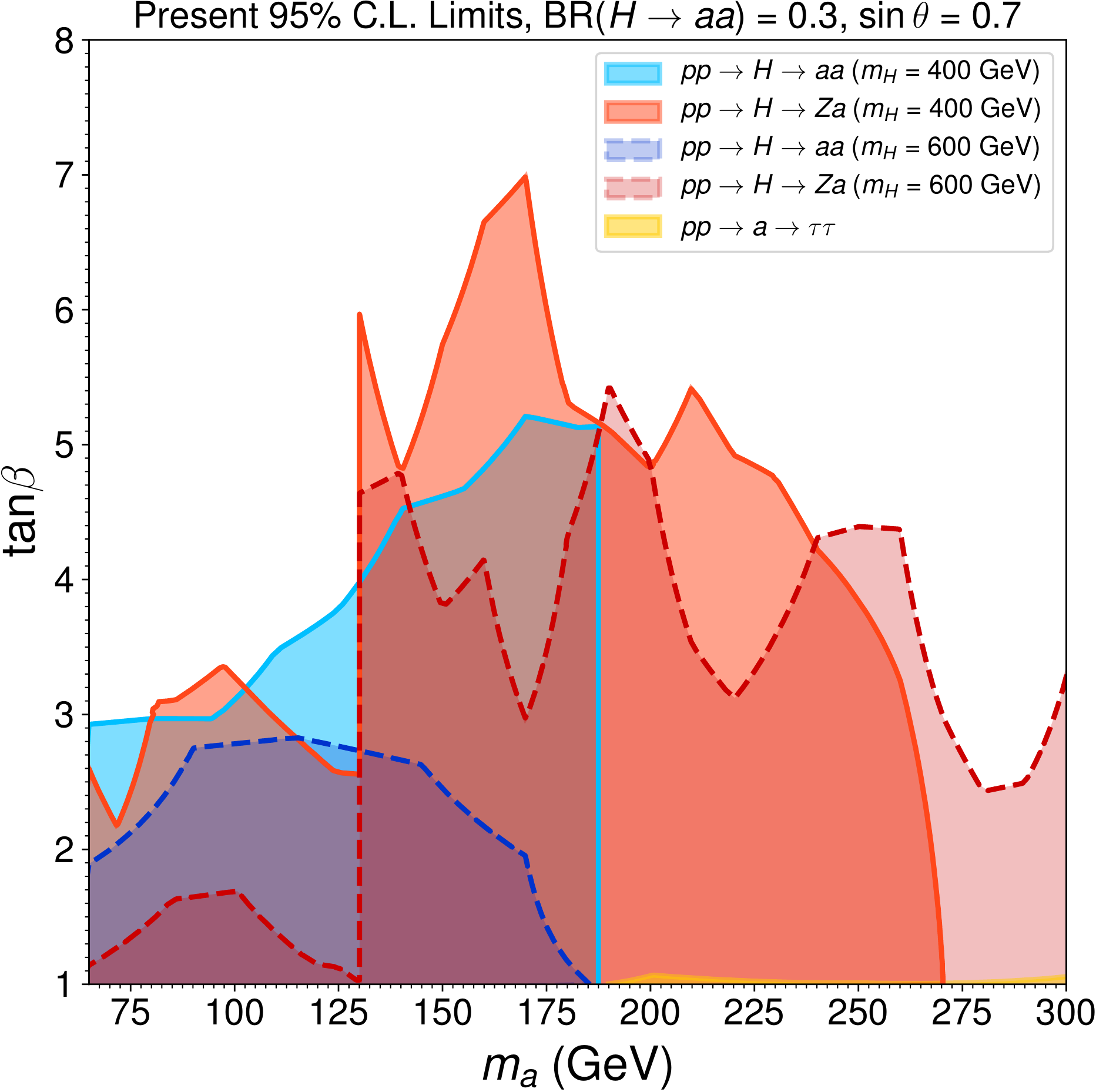}}}$
$\vcenter{\hbox{\includegraphics[width=0.495\textwidth]{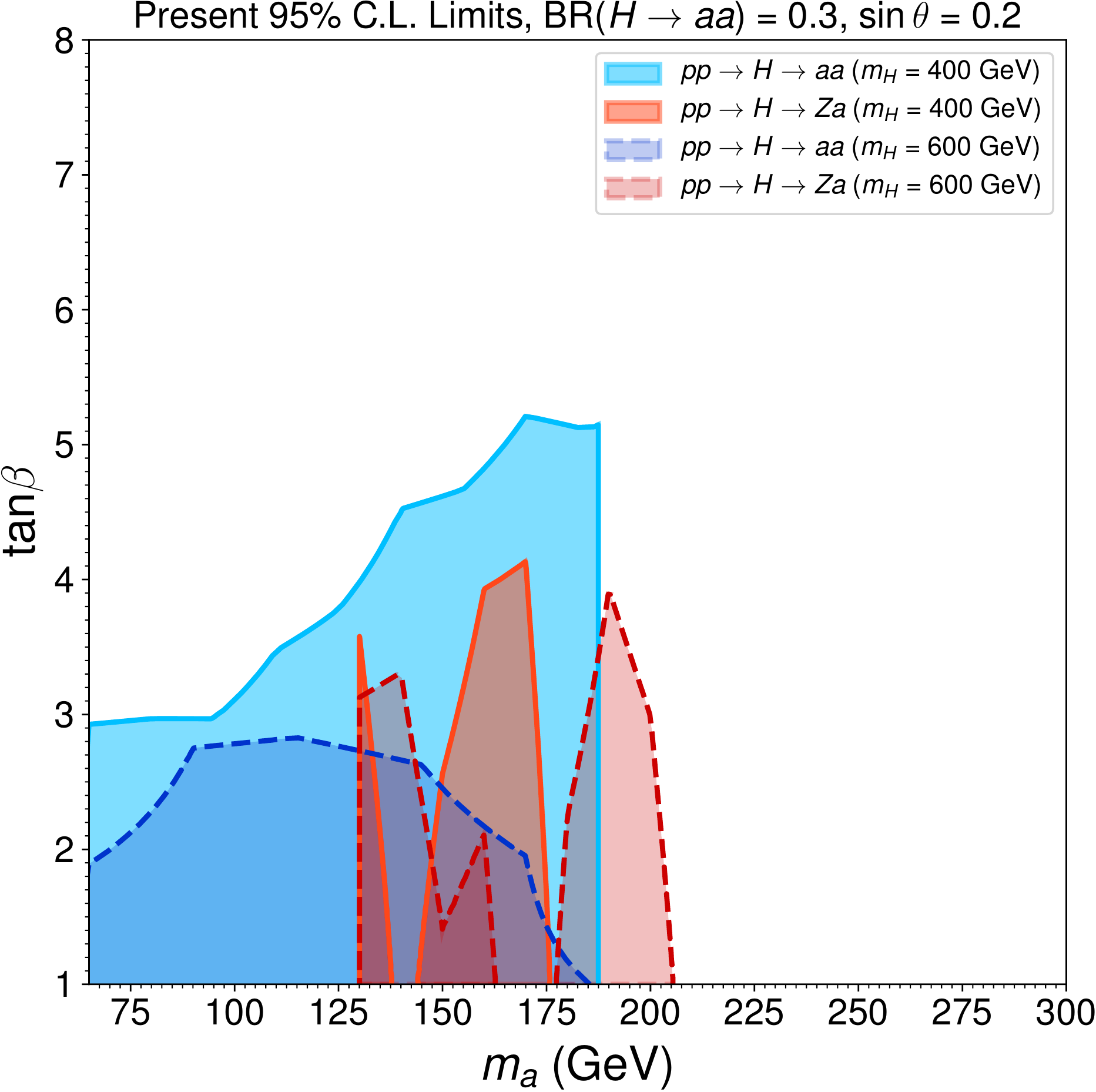}}}$

\caption{\small Present 95\% C.L. exclusion sensitivity/limits in the ($m_a$, tan$\,\beta$) plane for $c_{\beta-\alpha} = 0$, BR$(H \to aa) = 0.3$ and mixing $s_{\theta} = 0.7$ (left panel) and $s_{\theta} = 0.2$ (right panel), from $p p \to H \to a a \to b \bar{b} b \bar{b}$ searches (blue regions), $p p \to H \to Z a$ ($Z \to \ell\ell$, $a \to b \bar{b}$) ATLAS~\cite{Aaboud:2018eoy} and CMS~\cite{Khachatryan:2016are,CMS:2019wml} searches (red regions), and $p p \to a \to \tau \tau$ CMS~\cite{Sirunyan:2018zut} searches (yellow region), for $m_{H} = 400$ GeV (solid lines) and $m_{H} = 600$ GeV (dashed lines).}
\label{fig:2HDMS_Pseudoscalar}
\vspace{-2mm}
\end{figure}

We then perform a comparison of direct searches of $a$ via $p p \to a \to \tau \tau$ and $p p \to a \to \gamma \gamma$ with searches for $p p \to H \to Z a$ ($a \to b \bar{b}$) and our proposed search $p p \to H \to a a \to b \bar{b} b \bar{b}$, as a function of 
$m_H$, $m_a$, tan$\,\beta$, the singlet-doublet mixing $s_{\theta}$
and the branching fraction BR$(H \to aa)$. 
The various partial decay widths of $a$ are given in~\cite{Bauer:2017ota}, together with the partial decay widths of $H$ in the alignment limit $c_{\beta-\alpha} = 0$. The LHC production cross sections for $H$ and $a$ are computed at NNLO in QCD with \texttt{SusHi}~\cite{Harlander:2012pb,Harlander:2016hcx}.
Then, assuming the other 2HDM states ($H^{\pm}$ and $A$) are sufficiently heavy to not play a phenomenological role in the following, we show in Fig.~\ref{fig:2HDMS_Pseudoscalar} the present 95\% C.L. exclusion sensitivity in the ($m_a$, tan$\,\beta$) plane from our search 
$p p \to H \to a a \to b \bar{b} b \bar{b}$, for a fixed branching fraction BR$(H \to aa) = 0.3$ and mixing $s_{\theta} = 0.7$ (left panel) and $s_{\theta} = 0.2$ (right panel), considering $m_{H} = 400$ GeV (solid lines) and $m_{H} = 600$ GeV (dashed lines) as benchmarks. We also show the 
current 95\% C.L. exclusion limits from $p p \to H \to Z a$ ($Z \to \ell\ell$, $a \to b \bar{b}$) searches by ATLAS ($\sqrt{s} = 13$ TeV, 36.1 fb$^{-1}$~\cite{Aaboud:2018eoy}) and CMS ($\sqrt{s} = 8$ TeV, 19.8 fb$^{-1}$~\cite{Khachatryan:2016are} and $\sqrt{s} = 13$ TeV, 35.9 fb$^{-1}$~\cite{CMS:2019wml}), as well as from $p p \to a \to \tau \tau$ by CMS ($\sqrt{s} = 13$ TeV, 35.9 fb$^{-1}$~\cite{Sirunyan:2018zut}).   
We find $p p \to a \to \gamma \gamma$ searches (e.g.~\cite{CMS:2017yta}) are currently not sensitive to the parameter space of the model.

As highlighted in Fig.~\ref{fig:2HDMS_Pseudoscalar}, direct searches for $a$ barely have sensitivity to this model, with $p p \to a \to \tau \tau$ only constraining tan$\,\beta \lesssim 1$, and $p p \to a \to \gamma \gamma$ being even less sensitive at present. In contrast, searches for cascade scalar decays probe a sizable region of the parameter space, with a strong interplay between $H \to a a$ and $H \to Z a$ searches: for large singlet-doublet mixing (e.g.~left panel of Fig.~\ref{fig:2HDMS_Pseudoscalar}) $H \to Z a$ decays typically drive the sensitivity for $m_a > 130$ GeV, with $H \to a a$ providing the strongest sensitivity for $m_a < 130$ GeV (we note the ATLAS search~\cite{Aaboud:2018eoy} does not go below 130 GeV for the mass of the lighter BSM scalar); however, as the singlet-doublet mixing diminishes (right panel of Fig.~\ref{fig:2HDMS_Pseudoscalar}) the $H \to Z a$ sensitivity weakens significantly and the $H \to a a$ decay mode (which does not necessarily vanish in the limit $s_{\theta} \to 0$~\cite{Bauer:2017ota}) may become the leading probe of the parameter space of the model. 

The independence of the $H \to a a$ decay on the zero mixing limit $s_{\theta} \to 0$ is a strong point of this particular search mode. This is further emphasised by the fact that EW precision constraints from the $\rho$ parameter directly constrain $s_{\theta}$ as a function of the scalar sector masses~\cite{Bauer:2017ota}. For the 2HDM, the 
well-known BSM contributions to the $\rho$ parameter (see e.g.~\cite{Grimus:2008nb}) vanish for either $m_H=m_H^\pm$ or $m_A=m_H^\pm$ in the alignment limit. For non-zero $s_{\theta}$ the 2HDM pseudoscalar contribution is shared between the two CP-odd mass eigenstates, which yields an additional source of custodial symmetry breaking. In the decoupling regime for $A$ and $H^\pm$ considered here, the $\rho$ parameter drives the model towards either small singlet-doublet mixing or tuned regions of parameter space. Finally, let us stress that if the states $A$ and $H^{\pm}$ are not decoupled (contrary to what has been considered so far), other BSM scalar cascade decays not considered in this work such as $A \to h a$ and $H^{\pm} \to W^{\pm} a$ could allow to probe the 2HDM $+$ singlet pseudoscalar scenario. 

\vspace{2mm}

\noindent \underline{{\bf Scalar} $S$}: 

\vspace{1mm}

In this case the last three terms of~\eqref{2HDM_singlet_potential} may be present for a CP-conserving potential. After EW symmetry breaking, 
the singlet state $S$ (we consider for simplicity that the singlet field does not get a vev; the discussion when it does is however analogous) mixes with the CP-even states $H$ and $h$ from the 2HDM, yielding three CP-even mass eigenstates, one of which is the $125$ GeV Higgs boson. The other two states we label $H_1$ and $H_2$, with $m_{H_1} > m_{H_2}$. For simplicity, we focus on the limit where the $125$ GeV Higgs boson has SM-like properties\footnote{See~\cite{Baum:2018zhf,Carena:2015moc} for the alignment limit conditions in the 2HDM + $S$.}, and consider $H_2$ to be the singlet-like state. In this case, the decays of $H_2$ into SM particles are simply controlled by the mixing between the singlet $S$ and the heavy 2HDM state $H$ (see section~\ref{Sec:2HDM}), which we parametrise by sin$\,\theta$. Note that, contrary to the pseudoscalar case, the $\rho$ parameter does not necessarily imply constraints on this angle. These can be avoided in the alignment limit by having $m_A\simeq m_H^{\pm}$ and the cancellation is no longer spoiled by $a-A$ mixing.

\begin{figure}[h]

\centering

\includegraphics[width=0.8\textwidth]{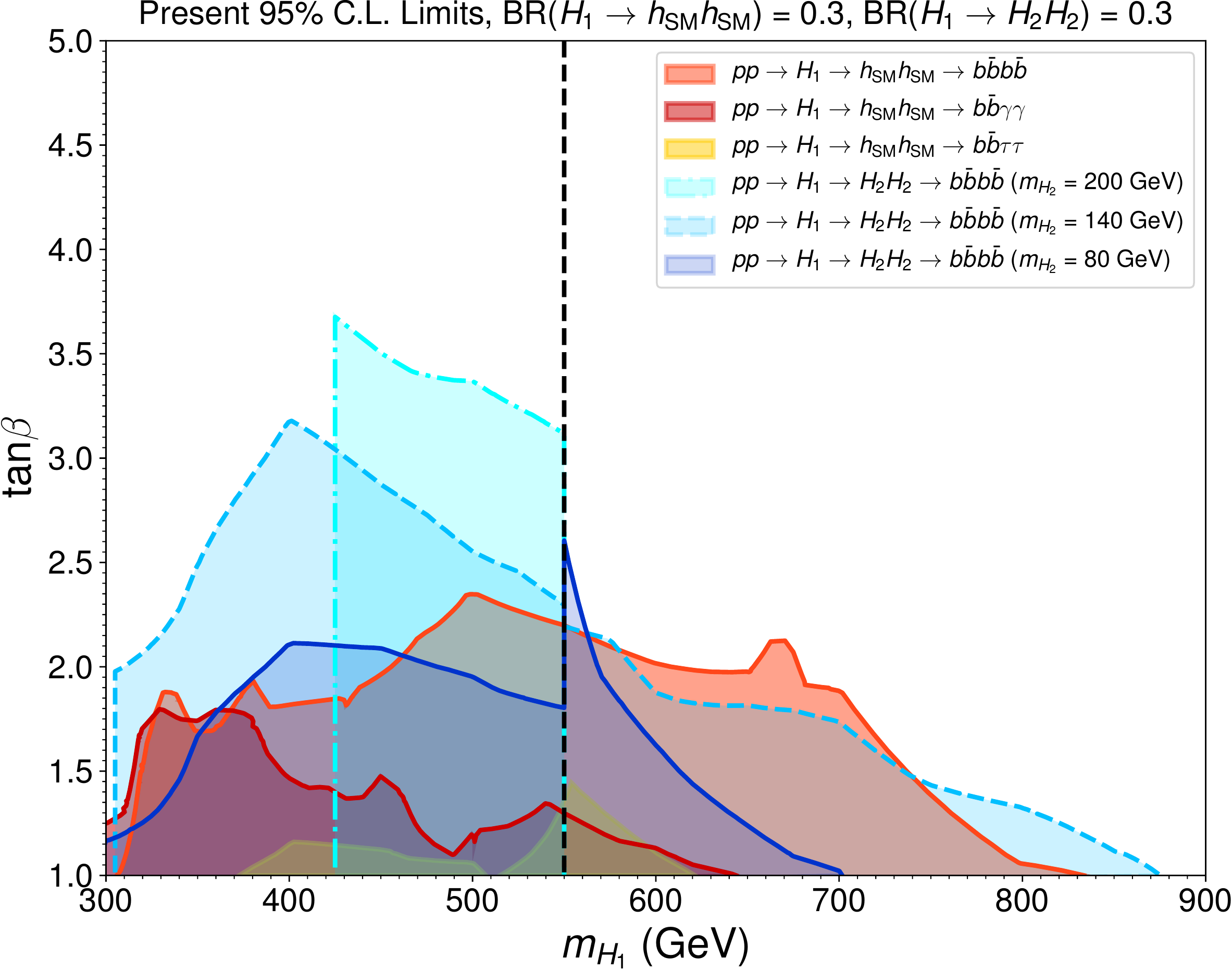}

\caption{\small Present 95\% C.L. exclusion sensitivity/limits in the ($m_{H_1}$, tan$\,\beta$) plane for fixed $c_{\beta-\alpha} = 0$, sin$\,\theta = 0.7$ and $\mathrm{BR}(H_1 \to h_{\mathrm{SM}} h_{\mathrm{SM}}) = \mathrm{BR}(H_1 \to H_2 H_2)  = 0.3$ from present ATLAS and CMS resonant di-Higgs searches $p p \to H_1 \to h_{\mathrm{SM}} h_{\mathrm{SM}}$ in the $b\bar{b} b \bar{b}$~\cite{Aaboud:2018knk,Sirunyan:2018zkk} (light red), 
$b\bar{b} \gamma\gamma$~\cite{Aaboud:2018ftw,Sirunyan:2018iwt} (dark red) and
$b\bar{b} \tau\tau$\cite{Aaboud:2018sfw,Sirunyan:2017djm} (yellow) final states, and from our proposed search $p p \to H_1 \to H_2 H_2 \to b \bar{b} b \bar{b}$ for $m_{H_2} = 80$ GeV (dark blue), $m_{H_2} = 140$ GeV (blue) and $m_{H_2} = 200$ GeV (light blue). The vertical dashed-black line corresponds to the boundary between the MMR ($m_{H_1} > 550$ GeV) and LMR ($m_{H_1} < 550$ GeV) categories of our search.
}
\label{fig:2HDMS_Scalar}

\vspace{-2mm}

\end{figure}

Again, we consider the case where the 2HDM states $A$ and $H^{\pm}$ are decoupled\footnote{If they are not, cascade decays such as $A \to Z H_2$ and $H^{\pm} \to W^{\pm} H_2$ could yield a probe of the 2HDM $+$ singlet scalar scenario.}. The leading decays of the state $H_1$ are $H_1 \to H_2 H_2$, $H_1 \to h_{\mathrm{SM}} h_{\mathrm{SM}}$ (this decay is possible in the alignment limit if $\mu_S$, $\mu_{S_1}$ and/or $\mu_{S_2}$ in~\eqref{2HDM_singlet_potential} are non-vanishing) and decays into SM fermions. At the same time, the LHC production cross section for $H_1$ is suppressed by cos$^2\theta$ compared to the production of the 2HDM state $H$ in the alignment limit $c_{\beta-\alpha} = 0$. In Fig.~\ref{fig:2HDMS_Scalar} we demonstrate the interplay between resonant di-Higgs searches $p p \to H_1 \to h_{\mathrm{SM}} h_{\mathrm{SM}}$ and our proposed search $p p \to H_1 \to H_2 H_2 \to b \bar{b} b \bar{b}$. Direct searches for the BSM state $H_2$ (e.g. $p p \to H_2 \to \tau\tau$ and $p p \to H_2 \to \gamma\gamma$) are currently only sensitive to the tan$\,\beta < 1$ and large mixing sin$\,\theta \to 1$ region of the 2HDM + scalar singlet scenario (similarly to what we already found above for the 2HDM + pseudoscalar singlet scenario).
Fig.~\ref{fig:2HDMS_Scalar} then shows, for fixed sin$\,\theta = 0.7$ and $\mathrm{BR}(H_1 \to h_{\mathrm{SM}} h_{\mathrm{SM}}) = \mathrm{BR}(H_1 \to H_2 H_2)  = 0.3$ the 95\% C.L. exclusion sensitivity in the ($m_{H_1}$, tan$\,\beta$) plane from present ATLAS and CMS resonant di-Higgs searches in the $b\bar{b} b \bar{b}$~\cite{Aaboud:2018knk,Sirunyan:2018zkk}, 
$b\bar{b} \gamma\gamma$~\cite{Aaboud:2018ftw,Sirunyan:2018iwt} and
$b\bar{b} \tau\tau$\cite{Aaboud:2018sfw,Sirunyan:2017djm} final states, and from our proposed search respectively for $m_{H_2} = 80$ GeV, 140 GeV and 200 GeV. The search $p p \to H_1 \to H_2 H_2 \to b \bar{b} b \bar{b}$ shows a comparable sensitivity to that of resonant di-Higgs for the whole range of $m_{H_2}$ for which the decay $H_1 \to H_2 H_2$ is open, and particularly in the LMR category it can probe significantly larger values of tan$\,\beta$.
This again highlights the potential role of this search as a discovery mode for non-minimal scalar sectors.

\section{Conclusions}
\label{sec:conclusions}

Searches for additional Higgs bosons at the LHC via new scalar decay modes are a key avenue to explore extensions of the SM Higgs sector. In this article we have presented the first study of the $p p \to H_1 \to H_2 H_2 \to b \bar{b} b \bar{b}$ channel, with both $H_1$ and $H_2$ being BSM states. A rather precise estimate of the LHC sensitivity of such a search is possible given its similarity with CMS and ATLAS resonant di-Higgs searches, which we have used to validate our analysis, specifically choosing for this purpose the latest $\sqrt{s} = 13$ TeV CMS resonant di-Higgs search 
in the $b \bar{b} b \bar{b}$ final state~\cite{Sirunyan:2018zkk}. The recasting procedure obtained here has the advantage of being model-independent, as it relies solely on the masses of the two BSM scalar particles. With present data from~\cite{Sirunyan:2018zkk}, the $p p \to H_1 \to H_2 H_2 \to b \bar{b} b \bar{b}$ search yields sensitivity to production cross sections times branching fractions ranging from the picobarn to tens of femtobarns depending on the BSM scalar masses, showing the power of this simple generalisation of an existing LHC search. We also stress that a dedicated experimental search is likely to yield appreciable improvements in sensitivity with respect to the one obtained in this work.

We have briefly discussed the impact of several experimental features that could affect the analysis, including $b$-tagging and the role of systematic errors. We have also devised a simplified procedure to obtain an estimate of the sensitivity for future collider machines, appropriately scaling the present LHC results to other center-of-mass energies and total integrated luminosities. We have analyzed the specific examples of the High Luminosity LHC (HL-LHC), the High Energy LHC (HE-LHC) and of the Future Circular Collider in its proton-proton incarnation (FCC-hh).

We have then applied our analysis to three specific scalar extensions of the SM: {\it i)} a Higgs sector with two additional singlet scalars; {\it ii)} a 2HDM scenario; {\it iii)} the 2HDM plus a singlet scalar/pseudoscalar, which is currently a ``de-facto" benchmark for dark matter searches at the LHC~\cite{Abe:2018bpo}. Comparing the reach of our proposed study to present LHC searches constraining these models, we explicitly show the parameter space regions that our search renders accessible, stressing its complementarity to existing searches. 

Our study shows promising prospects for this yet unexplored probe of heavy Higgs bosons, and highlights explicitly how extending the coverage of current LHC searches for BSM scalars can yield new avenues to probe non-minimal Higgs sectors. 
Finally, we note that our study represents the first, minimal step in probing scenarios with Higgs-to-Higgs decay topology in which all states come from the BSM sector. As the paradigm for extended scalar sectors evolves into increasingly non-minimal territory, it is essential that the experimental programme continues to extend its searches to probe uncharted model space through Higgs-to-Higgs cascades.

\section*{Acknowledgements}
J.Z. is indebted to Bernhard Mistlberger, Stefan Liebler and Tania Robens for useful discussions. K.M. is supported in part by the F.R.S.-FNRS under the Excellence of Science EOS be.h project n. 30820817 and by the European Union's Horizon 2020 research and innovation programme
under the Marie Sklodowska-Curie grant agreement No. 707983. K.M. would like to thank Olivier Mattelaer for valuable advice regarding the signal generation.
J.M.N. was partially supported by the Programa Atracci\'on de Talento de la Comunidad de Madrid under grant 2017-T1/TIC-5202, and by Ram\'on y Cajal Fellowship contract RYC-2017-22986.
J.M.N also acknowledges support from the Spanish MINECO's ``Centro de Excelencia Severo Ochoa" Programme under grant SEV-2016-0597, from the European Union's Horizon 2020 research and innovation programme under the Marie Sklodowska-Curie grant agreements 690575  (RISE InvisiblesPlus) and 674896 (ITN ELUSIVES) and from the Spanish Proyectos de I$+$D de Generaci\'on de Conocimiento via grant PGC2018-096646-A-I00.
C.V. is supported by the SLAC Panofsky Fellowship.
D.B. thanks the Galileo Galilei Institute for theoretical physics 
for hospitality while part of this work was carried out.
J.M.N. thanks the Korean Institute for Advanced Study (KIAS) for hospitality during the last stages of this work.

\begin{appendix}
\section{Improved $b$-tagging parametrization}
\label{app:btagging}
\setcounter{equation}{0}
\renewcommand{\theequation}{A.\arabic{equation}}

Given the 4 $b$-quark final state of our analysis, our results are dependent on a good modeling of $b$-tagging performance. 
{\tt Delphes}~\cite{deFavereau:2013fsa} is used throughout our phenomenological analysis, which admits $b$-tagging efficiencies as a function of jet kinematics. In order to replicate as closely as possible the behaviour of the {\tt DeepCSV} medium $b$-tagging working point employed in the CMS search, we implemented a {\tt Delphes} card that parametrised the tagging efficiency using the information reported in \cite{Sirunyan:2017ezt}. In that study, $b$-tagging efficiencies and $c$- and light-jet mis-tag rates are determined using a high purity $t\bar{t}$ sample and quoted in bins of either $p_T$ or $\eta$, but not both simultaneously. 1D parametrisations as a function of $p_T$ are reported in several $p_T$ bins for the three working points. The medium working point has an inclusive $b$-tagging efficiency of 68\%, and inclusive mis-tag rates of 12\% and 1.1\% for $c$- and light-jets, respectively.  In order to have a better modelling of the $b$-tagging efficiency over the jet kinematics, we extrapolate the reported efficiencies into a 2D function of jet $p_T$ and $\eta$. 

\begin{figure}[ht!]
\begin{center}
\includegraphics[width=0.75\textwidth]{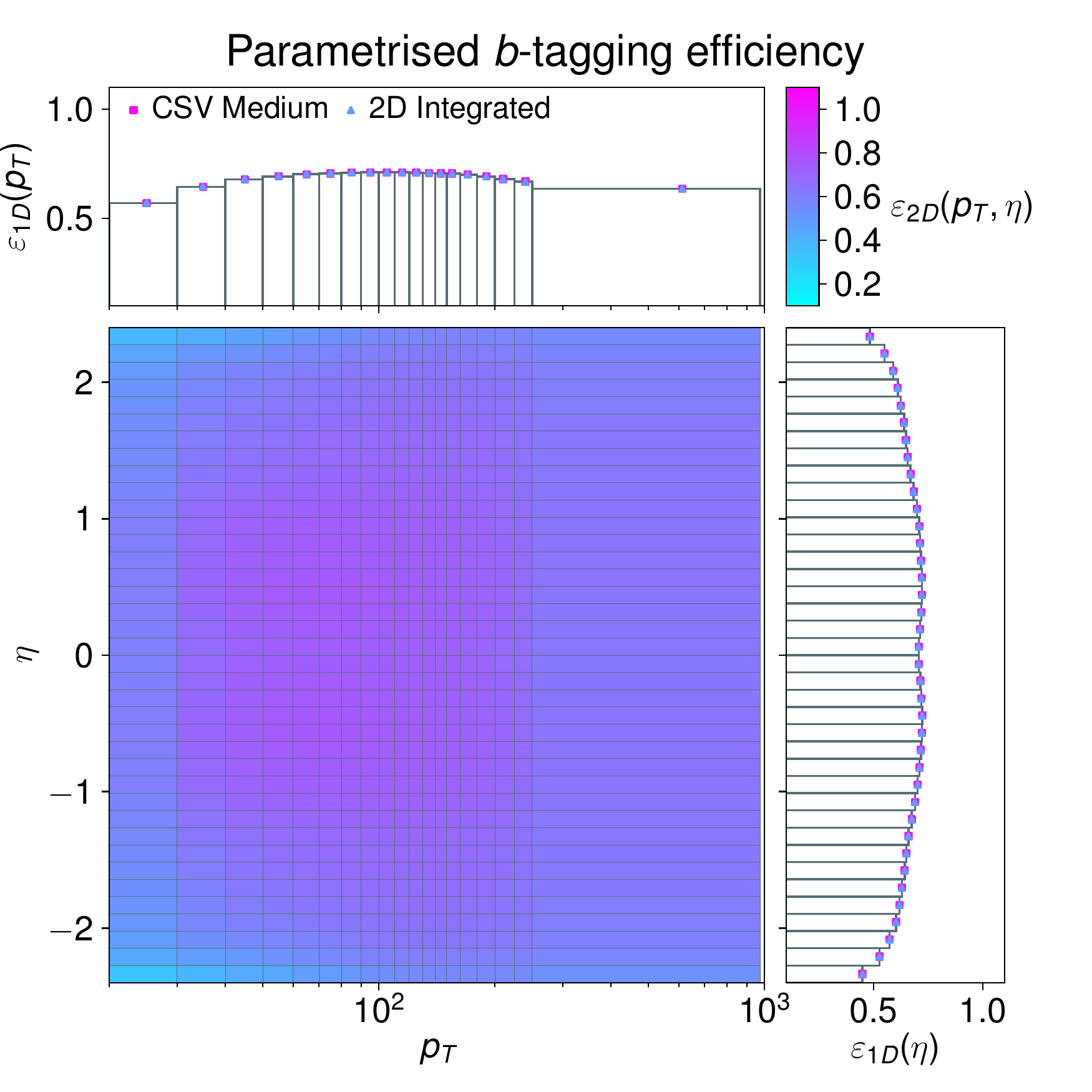}

\caption{\small 2-dimensional interpolated $b$-tagging efficiency map obtained from the information in~\cite{Sirunyan:2017ezt}. The upper and right panel show the integrated efficiencies in $p_T$ and $\eta$, obtained by convoluting the 2D efficiency map with a double-differential distribution of $b$-jets from $t\bar{t}$ (blue triangles). These are shown alongside the 1D efficiency maps reported in~\cite{Sirunyan:2017ezt} (pink squares). \label{Btag_HAA}}
\end{center}

\vspace{-2mm}

\end{figure}

Our $b$-tagging efficiency map, shown in Fig.~\ref{Btag_HAA}, is determined by a fit to these results using the reported 1D efficiencies as boundary conditions, taking into account the kinematical distributions of $b$-jets from $t\bar{t}$. A $t\bar{t}$ sample was generated at NLO in QCD with {\tt Madgraph5\_aMC@NLO}, to obtain a 2-dimensional distribution in $p_T$ and $\eta$ with binnings that matched the reported efficiency curves of \cite{Sirunyan:2017ezt}. The `unfolded' 2D efficiency map must obey the constraint that it reproduces the reported 1D maps when integrated along either axis as well as the inclusive efficiency when fully integrated over. The integration procedure is a convolution of the binned efficiency map with the double-differential distribution in $p_T$ and $\eta$.  
As shown in the upper and right panels of Fig.~\ref{Btag_HAA}, the solution is able to satisfy the constraints (the last 10 $p_T$ bins were combined due to lack of MC statistics in the $t\bar{t}$ sample). However, this is clearly an under-constrained problem, and the solutions are found to be somewhat sensitive to the initial conditions of the least-squares minimisation procedure used. The map shown in Fig.~\ref{Btag_HAA} used the average of the corresponding efficiencies in the 1D $p_T$ and $\eta$ maps as a starting point. Taking randomised initial conditions leads to noisy solutions that oscillate around the former. This was verified by averaging over a stochastic sample of solutions with random initial conditions between 0 and 1, observing that the resulting map was within 10--20\% of that obtained from the average initial conditions. The variance of the obtained efficiency in the regions where most of the $t\bar{t}$ sample resided was found to be around 10--20\%. This region has a dominant impact on the integrated efficiencies. In bins poorly populated by $t\bar{t}$, the results fluctuated more, with a standard deviation of order 50\%. These bins, however, do not have a big impact of the overall efficiency.

For the $c$- and light-jet mis-tag rates, a simpler, more approximate procedure was employed. The 1D efficiencies in $p_T$ and $\eta$ were taken as independent and used to construct a 2D map, for each $p_T$ bin where the polynomial parametrisations were provided by the CMS collaboration. In each bin, the efficiency was taken as the product of the $p_T$-dependent parametrisation and a fit to the (inclusive) $\eta$ efficiency distribution of the medium working point, divided by the integral of said $\eta$ distribution, such that the integral of the new map matched the integral of the 1D $p_T$ parametrisation. These rates are not expected to have any significant impact on our analysis as the multi-jet background is determined by data-driven methods. They therefore did not warrant a high-statistics MC simulation of the non $b$-jet composition/kinematics in $t\bar{t}$ that would be required to repeat the procedure employed for the $b$-tagging. 

\begin{figure}[h!]
\centering

\includegraphics[width=0.99\textwidth]{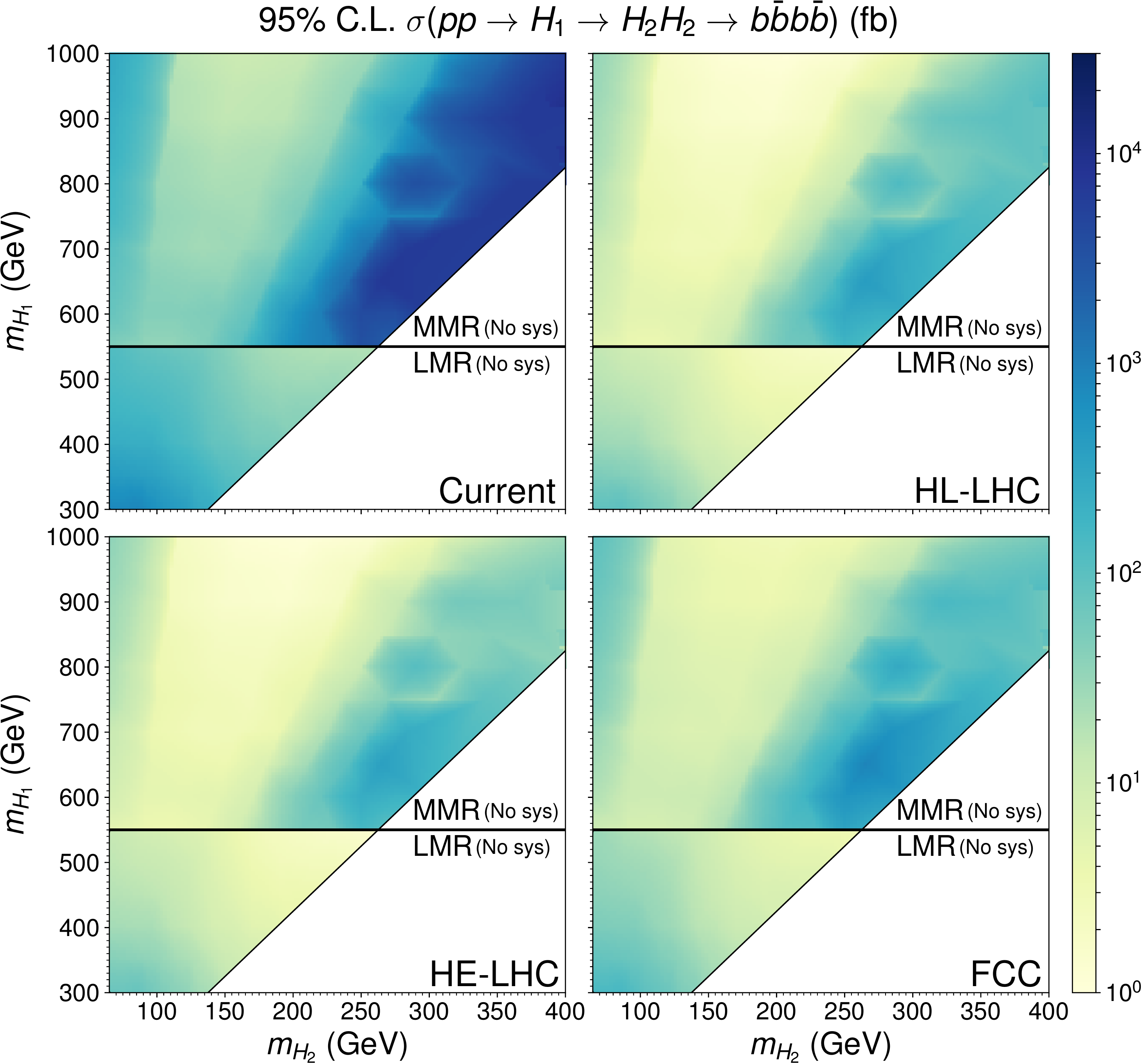}

\caption{\small 95\% C.L. upper limit on the $pp\to H_1 \to H_2 H_2 \to b\bar b b \bar b$ signal cross section (in fb) in the ($m_{H_2}, \, m_{H_1}$) 
plane for the MMR ($m_{H_1} > 550$ GeV) and LMR ($m_{H_1} < 550$ GeV) categories, with \textbf{no SM background systematic uncertainties} accounted for. Top-left, top-right, bottom-left and bottom-right panels correspond respectively to current LHC data ($\sqrt{s} = 13$ TeV, 35.9 fb$^{-1}$) HL-LHC, HE-LHC and FCC-hh.}
\label{fig:AppendixB_Limits}
\end{figure}

\section{$H_1 \to H_2 H_2 \to b \bar{b} b \bar{b} $ limits/projections without systematic uncertainties}
\label{app:H1H2H2nosys}
\setcounter{equation}{0}
\renewcommand{\theequation}{B.\arabic{equation}}

Here we provide our current ($\sqrt{s} = 13$ TeV, 35.9 fb$^{-1}$) LHC estimates and future HL-LHC, HE-LHC and FCC-hh projections for the 95\% C.L. cross section sensitivity for the process $p p \to H_1 \to H_2 H_2 \to b \bar{b} b \bar{b}$ assuming no systematic uncertainties, illustrated in Fig.~\ref{fig:AppendixB_Limits}. In addition, we give the values of $\kappa^2 \times \mathrm{BR}$ and the improvement in sensitivity $I$ (for HL-LHC, HE-LHC and FCC-hh with respect to the current sensitivity) for the ($m_{H_2}$, $m_{H_1}$) plane benchmarks defined previously in Table~\ref{Table2}, in the absence of SM background systematic uncertainties, in  Table~\ref{Table3}.

\begin{table}[h!]
\begin{center}
\begin{tabular}{|C{0.6cm}|C{0.6cm}|C{2.5cm}|C{1.7cm}|C{0.75cm}|C{1.7cm}|C{0.75cm}|C{1.7cm}|C{0.75cm}|}
\cline{3-9}
\multicolumn{2}{c|}{ } & LHC 35.9 fb$^{-1}$  & \multicolumn{2}{c|}{HL-LHC}  & \multicolumn{2}{c|}{HE-LHC} & \multicolumn{2}{c|}{FCC-hh}  \\ \hline
$m_{H_2}$ & $m_{H_1}$ & $\kappa\times$BR & $\kappa\times$BR & $I$ & $\kappa\times$BR & $I$ & $\kappa\times$BR & $I$  \\ \hline
\multirow{4}{*}{75}
&  300  & $7.0\times 10^{-2}$ & $7.1\times 10^{-3}$ &  9.9   & $1.7\times 10^{-3}$ &  40.3  & $4.5\times 10^{-4}$ &  155 \\ \cline{2-9}
&  500  & $2.7\times 10^{-2}$ & $2.7\times 10^{-3}$ &  10.0  & $6.1\times 10^{-4}$ &  44.8  & $1.4\times 10^{-4}$ &  193 \\ \cline{2-9}
&  700  & $8.4\times 10^{-2}$ & $8.2\times 10^{-3}$ &  10.3  & $1.7\times 10^{-3}$ &  49.9  & $3.5\times 10^{-4}$ &  237 \\ \cline{2-9}
&  900  & $7.4\times 10^{-1}$ & $7.2\times 10^{-2}$ &  10.4  & $1.4\times 10^{-2}$ &  54.5   & $2.6\times 10^{-3}$ &  282 \\ \hline
\multirow{4}{*}{125}
&  300  & $3.0\times 10^{-2}$ & $3.0\times 10^{-3}$ &  10.0  & $7.4\times 10^{-4}$ &  40.6  & $1.9\times 10^{-4}$ &  156  \\ \cline{2-9}
&  500  & $1.8\times 10^{-2}$ & $1.8\times 10^{-3}$ &  10.2  & $4.0\times 10^{-4}$ &  45.5  & $9.2\times 10^{-5}$ &  196 \\ \cline{2-9}
&  700  & $2.7\times 10^{-2}$ & $2.5\times 10^{-3}$ &  10.7  & $5.2\times 10^{-4}$ &  52.2  & $1.1\times 10^{-4}$ &  248 \\ \cline{2-9}
&  900  & $7.4\times 10^{-2}$ & $6.7\times 10^{-3}$ &  11.0  & $1.3\times 10^{-3}$ &  57.8  & $2.5\times 10^{-4}$ &  300 \\ \hline
\multirow{3}{*}{175}
&  500  & $6.7\times 10^{-3}$ & $6.2\times 10^{-4}$ &  10.7  & $1.4\times 10^{-4}$ &  48.1  & $3.2\times 10^{-5}$ &  207 \\ \cline{2-9}
&  700  & $3.7\times 10^{-2}$ & $2.9\times 10^{-3}$ &  12.5   & $5.9\times 10^{-4}$ &  61.7  & $1.2\times 10^{-4}$ &  294 \\ \cline{2-9}
&  900  & $6.5\times 10^{-2}$ & $4.9\times 10^{-3}$ &  13.1  & $9.2\times 10^{-4}$ &  70.1  & $1.8\times 10^{-4}$ &  364 \\ \hline
\multirow{3}{*}{225}
&  500  & $5.9\times 10^{-3}$ & $4.9\times 10^{-4}$ &  12.0  & $1.1\times 10^{-4}$ &  54.5   & $2.5\times 10^{-5}$ &  235 \\ \cline{2-9}
&  700  & $2.8\times 10^{-1}$ & $1.8\times 10^{-2}$ &  15.6  & $3.5\times 10^{-3}$ &  78.6  & $7.4\times 10^{-4}$ &  376 \\ \cline{2-9}
&  900  & $1.2\times 10^{-1}$ & $6.4\times 10^{-3}$ &  18.5  & $1.2\times 10^{-3}$ &  102  & $2.2\times 10^{-4}$ &  535 \\ \hline
\multirow{2}{*}{275}
&  700  &  $5.0\times 10^{0\phantom{-}}$   & $1.8\times 10^{-1}$ &  27.5  & $3.4\times 10^{-2}$ &  149  & $7.0\times 10^{-3}$ &  723  \\ \cline{2-9}
&  900  &  $1.3\times 10^{0\phantom{-}}$    & $4.3\times 10^{-2}$ &  29.9  & $7.3\times 10^{-3}$ &  177  & $1.4\times 10^{-3}$ &  936 \\ \hline
\multirow{2}{*}{325}
&  700  &  $6.4\times 10^{0\phantom{-}}$  & $1.4\times 10^{-1}$ &  45.2  & $2.3\times 10^{-2}$ &  278  & $4.6\times 10^{-3}$ &  1399 \\ \cline{2-9}
&  900  &  $1.6\times 10^{1\phantom{-}}$ & $2.6\times 10^{-1}$ &  59.3  & $3.6\times 10^{-2}$ &  435  & $6.5\times 10^{-3}$ &  2407 \\ \hline
\end{tabular}
\caption{\small Value of $\kappa^2 \times$BR for different $m_{H_2}$, $m_{H_1}$ (in GeV) benchmarks, for our proposed search, with \textbf{no SM background systematic uncertainties} accounted for. Top-left, top-right, bottom-left and bottom-right panels correspond respectively to current LHC data ($\sqrt{s} = 13$ TeV, 35.9 fb$^{-1}$) HL-LHC, HE-LHC and FCC-hh. We also give the sensitivity improvement $I$ for HL-LHC, HE-LHC and FCC-hh, in all cases with no SM background systematic uncertainties included.} 
\label{Table3}
\end{center}

\vspace{-3mm}

\end{table}

\end{appendix}

\bibliographystyle{JHEP}
\bibliography{HAA_BIB}

\end{document}